\documentclass[notitlepage,twocolumn,prb,aps,showpacs,10pt,superscriptaddress]{revtex4-1}

\usepackage{amsmath}
\usepackage{graphicx}
\usepackage{xcolor}
\usepackage{footnote}
\usepackage{textcomp}
\usepackage{hyperref}
\hypersetup{colorlinks=true,allcolors=blue}

\newcommand{\mbf}[1]{\mathbf{#1}}
\renewcommand{\t}[1]{\textrm{#1}}
\newcommand{\nn}{\nonumber\\}

\newcommand{\q}{\mbf{q}}

\newcommand{\g}{\gamma}

\renewcommand{\k}{\kappa}

\newcommand{\m}{\mu}
\newcommand{\n}{\nu}
\newcommand{\x}{\xi}
\renewcommand{\r}{\rho}
\newcommand{\s}{\sigma}

\newcommand{\w}{\omega}

\newcommand{\D}{\Delta}

\newcommand{\barr}{\bar{\r}}
\newcommand{\dbarr}{\bar{\barr}}

\newcommand{\+}{^\dagger}
\renewcommand{\>}{\rangle}
\newcommand{\<}{\langle}
\newcommand{\Tr}{\t{Tr}}

\newcommand{\G}{\vert G\>}
\newcommand{\X}{\vert X\>}

\newcommand{\XX}{\vert X\>\<X\vert}
\newcommand{\XG}{\vert X\>\<G\vert}
\newcommand{\GX}{\vert G\>\<X\vert}

\begin{document}
\title{Path-integral approach for nonequilibrium multi-time correlation functions of open quantum systems
coupled to Markovian and non-Markovian environments}
\author{M. Cosacchi}
\affiliation{Theoretische Physik III, Universit{\"a}t Bayreuth, 95440 Bayreuth, Germany}
\author{M. Cygorek}
\affiliation{Department of Physics, University of Ottawa, Ottawa, Ontario, Canada K1N 6N5}
\author{F. Ungar}
\affiliation{Theoretische Physik III, Universit{\"a}t Bayreuth, 95440 Bayreuth, Germany}
\author{A. M. Barth}
\affiliation{Theoretische Physik III, Universit{\"a}t Bayreuth, 95440 Bayreuth, Germany}
\author{A. Vagov}
\affiliation{Theoretische Physik III, Universit{\"a}t Bayreuth, 95440 Bayreuth, Germany}
\affiliation{ITMO University, 49 Kronverksky Pr., St. Petersburg, 197101, Russia}
\author{V. M. Axt}
\affiliation{Theoretische Physik III, Universit{\"a}t Bayreuth, 95440 Bayreuth, Germany}
\begin{abstract}
Using a real-time path integral approach we develop an algorithm to calculate multi-time
correlation functions of open few-level quantum systems that is applicable to highly
nonequilibrium dynamics.
The calculational scheme fully keeps the non-Markovian memory introduced by the 
pure-dephasing type coupling to a continuum of oscillators.
Furthermore, we discuss how to deal consistently with the simultaneous presence of 
non-Markovian and Markovian system reservoir interactions.
We apply the method to a crucial test case, namely the evaluation of
emission spectra of a laser-driven two-level quantum dot  coupled to a continuum of 
longitudinal acoustic phonons, which give  rise to non-Markovian dynamics.
Here, we also account for the coupling to a photonic environment, which models radiative
decay and can be treated as a Markovian bath.
The phonon side bands are found on the correct side of the zero phonon line in our calculation, in contrast
to known results where the quantum regression theorem is applied naively to
non-Markovian dynamics.
Combining our algorithm with a recently improved iteration scheme for performing the
required sum over paths we demonstrate the numerical feasibility of our approach to
systems with more than two levels.
Results are shown for the second-order photonic two-time correlation function of 
a quantum dot-cavity system with seven states on the
Jaynes-Cummings ladder taken into account.
\end{abstract}
\maketitle

\section{Introduction}

To understand the dynamics of quantum few-level systems influenced by a huge number 
of degrees of freedom that one commonly refers to as a bath is of major importance in 
many areas of physics. Often, matters are further complicated by the fact that baths 
of many types couple simultaneously to the subsystem of interest.
For example, solid state quantum emitters can couple to photonic reservoirs, i.e., free
electromagnetic field modes as well 
as to phonons that are always present in a solid state environment.
While the former can usually be approximated in a Markovian fashion to a reasonable 
extent, the latter is known to show pronounced non-Markovian behavior.
\cite{Foerstner2003,Vagov2004,Thorwart2005,Vagov2007,Ramsay2010a,Kaer2010}
In the case of strongly confined quantum dots that have been proposed as candidates 
for numerous technological
applications in quantum communications, quantum computing, and quantum cryptography 
as sources of single-photons\cite{Michler2000,He2013,Somaschi2016} and entangled 
photon pairs\cite{Stevenson2006,Akopian2006,Hafenbrak2007,Dousse2010,Mueller2014}, 
the coupling of longitudinal acoustic phonons to the system is usually modeled as 
being of the pure-dephasing type.\cite{Besombes2001,Borri2001,Krummheuer2002,Ramsay2010a}
The non-Markovian nature of this interaction manifests itself in a finite but not 
necessarily short memory for the reduced system density matrix.

Multi-time correlation functions contain a huge amount of important information about
open quantum systems. They constitute the interface between theory and 
a large class of measurements. Emission spectra that can easily be obtained in 
experiment are represented by the Fourier transform of the first-order two-time 
photon correlation function.
\cite{Besombes2001,Ulrich2011,Roy2011a,Roy2011b,Harsij2012,Roy2012,McCutcheon2013,
Roy-Choudhury2015,McCutcheon2016}
In the stationary emission spectrum of a quantum dot for example, it can clearly
be seen that the inclusion of the non-Markovian phonon dynamics is essential since
it manifests itself in distinct features like an energetically broad phonon
sideband at low temperatures.\cite{McCutcheon2016}
Multi-time correlation functions with two or more time arguments can be recorded 
in coincidence measurements.\cite{Hanbury1956,Glauber2007}
Such experiments shed light on photonic properties of the system like, e.g., (anti-)
bunching, photon purities or indistinguishabilities.\cite{Santori2002,Bentham2016,Prtljaga2016,Weiss2016,Ding2016,Schweickert2017}
Although the most commonly considered case is the two-time correlation function,
multi-time correlation functions with more than two time arguments can also be a valuable
tool, e.g., for the characterization of the statistics of photon bundles.\cite{Munoz2014}

The theoretical analysis of multi-time correlations is a challenging task, especially in systems
with extended memory.
The most widely used approximate approach is 
is based on the quantum regression theorem (QRT).
In the derivation of the QRT it is assumed that the statistical operator factorizes at all times into 
a system and an environment part.\cite{Carmichael1993}
In addition, the environment-induced dynamics is assumed to be Markovian.
With these assumptions the two-time correlation function is calculated in two steps:
(i) one obtains the dependence of the reduced density matrix on the first time argument $t$
by solving the master equation within the Markov approximation and (ii) 
Then the result of the $t$-propagation is used to construct suitable initial values for 
the subsequent propagation in the second argument $\tau$, which relies on the same 
dynamical equation.
It is tempting to extend this QRT-based approach to treat a system with a non-Markovian environment
simply by replacing the Markovian propagator for the $t$- and $\tau$-evolution with the
non-Markovian one. 
Such a naive extension, however, completely neglects the influence of the memory built up during
the $t$-propagation on the subsequent $\tau$-evolution since this memory
is discarded by starting the $\tau$- propagation as an initial 
value problem.
Indeed, it has been recently demonstrated\cite{McCutcheon2016} that this approach may lead to unphysical
results. In particular, it predicts phonon sidebands in the emission spectra on the wrong side of the
Mollow triplet.
In Ref.~\onlinecite{McCutcheon2016} this problem has been overcome by accounting 
perturbatively for corrections to the QRT.

Furthermore, the system may couple to Markovian baths simultaneously to the above 
described interaction with a non-Markovian environment.
When a bath is known to influence a system only in Markovian ways, it is 
unnecessary and typically also impractical to account for the system-bath coupling 
by formulating the Hamiltonian time evolution for the total system.
One usually approximates these bath influences by inserting Lindblad-type 
superoperators into the equation of motion for the system's reduced density 
matrix. Therefore, the reduced Markovian dynamics becomes non-Hamiltonian and 
thus non-unitary.
However, the definition of multi-time correlation functions makes use of the 
Heisenberg picture in order to assign the time arguments to the operators of interest, 
i.e., it relies on the existence of a unitary time evolution operator.
If all baths are Markovian, one can use the QRT and ends up with Lindblad-type 
contributions for both the $t$- and the $\tau$-propagation. So the question arises 
what to do when the QRT cannot be used because of the memory induced by a non-Markovian 
environment.

In the present paper we present a practical scheme for evaluating multi-time correlation 
functions based on path integral (PI) methods.
\cite{Makri1995a,Makri1995b,Thorwart2000,Vagov2011}
Most importantly, the scheme fully accounts for the non-Markovian memory induced in
the system by the pure-dephasing type coupling to a continuum of harmonic
oscillators.
This allows the inclusion of the memory built up during the $t$-propagation and start
the $\tau$-propagation without assuming an initial value problem, contrary to the QRT
approach.
In addition, it is shown how to consistently include the impact of non-Hamiltonian 
Markovian contributions to the dynamics, while keeping the accuracy with respect to 
the non-Markovian environment.
Our approach is a considerable extension of  earlier PI-based calculations\cite{Shao2002}
which addressed equilibrium two-time correlation 
functions that depend only on the time difference
$\tau$ and which did not consider 
additional Markovian baths.
It also extends 
PI descriptions where simultaneously Hamiltonian and 
non-Hamiltonian contributions have been treated for the evaluation of single-time 
expectation values\cite{Barth2016} to the case of multi-time functions.
Finally, our algorithm can be combined with a recently developed improved
iteration scheme for performing the required sum of paths\cite{Cygorek2017}
which greatly enhances the numerical efficiency such that quantum systems
with much more than two levels can be explored.
Moreover, we would like to note that recently an article has been brought to our attention in which the path integral algorithm was optimized even further by employing tensor-network techniques. \cite{Strathearn2018}
Combining this approach with the algorithm presented here may be used to greatly extend the range of applicability of PI methods.

We also would like to mention 
that a first application of our algorithm beyond the proof of principle
calculations shown in the present paper can be found in
Ref.~\onlinecite{Cygorek2017b},
where the second-order two-time correlation function $G^{(2)}(t,\tau)$
has been evaluated for an exciton-biexciton QD system coupled
to two cavity modes.
Ref.~\onlinecite{Cygorek2017b} gave only the final results without any
description of the used formalism.  Here we fill the gap by presenting details
of the method in a general formulation that allows one to calculate arbitrary
multi-time correlations.  As illustrative examples we present results for the
two-time correlation function in the photon anti-bunching regime and for emission
spectra demonstrating that our algorithm yields phonon sidebands energetically
on the correct side.

\section{Model}
\label{sec:Model}

To demonstrate the derivation of our algorithm, we choose the probably simplest prototype
of an open quantum dissipative system coupled simultaneously
to a non-Markovian and a Markovian environment exhibiting a non-stationary driven
dynamics, which is provided by an externally driven two-level system with pure-dephasing coupling
to an oscillator continuum and an additional reservoir.
Since all formal developments of the present paper can be discussed within this
showcase model we shall present all derivations explicitly for this case, also
to keep the notation simple.
Generalizations, e.g., to systems with more levels or further bath couplings are straightforward.

To be specific, we study a strongly confined self-assembled two-level quantum dot (QD)
driven by an external laser field and coupled to a continuum of longitudinal acoustic (LA) phonons
as well as to a photonic reservoir.
The phonon coupling is commonly assumed to be
of pure-dephasing type and is known to give rise
to non-Markovian features in many cases.
\cite{Foerstner2003,Vagov2004,Thorwart2005,Vagov2007,Ramsay2010a,Kaer2010}
The interaction with a broad photonic reservoir is the origin of radiative decay
which is typically to a good approximation represented as a Markovian process.

The Hamiltonian for this system is given by
\begin{subequations}
\begin{align}
\label{eq:Hamiltonian_model}  
H(t)&= H_{1}(t) + H_{\t{Rad}}+H_{\t{QD-Rad}}\, ,\\
H_{1}(t)&= H_{0}(t) +H_{\t{Ph}}+H_{\t{QD-Ph}}\, ,\\
H_{0}(t)&=H_{\t{QD}}+H_{\t{driv}}(t)\, ,
\end{align}
\end{subequations}
where $H_{0}(t)$ represents the laser driven QD without reservoir coupling
which is actually the subsystem of interest.
$H_{1}(t)$ in addition comprises the interaction with the phonons and, finally
$H(t)$ is the total Hamiltonian where also the coupling to the photonic
reservoir is accounted for.
In the common dipole and rotating wave approximation in a frame co-rotating with
the laser frequency, the two-level and the driving parts of the Hamiltonian are
given by 
\begin{align}
H_{\t{QD}}&=\hbar\D\w_{\t{XL}}\XX\\ 
\text{and}\quad
H_{\t{driv}}(t)&=-\frac{\hbar}{2}\left(f(t)\XG+f^*(t)\GX\right)\, ,
\end{align}
respectively, where $\G$ denotes the ground and $\X$ the excited state of the dot.
$f(t)$ comprises the envelope of the driving
laser field that is detuned by $\D\w_{\t{XL}}$ from the ground to
excited state transition frequency of the dot
as well as the dipole matrix element for the transition.
Note that $H_{\t{driv}}(t)$ introduces an explicit time-dependence and puts the system out
of equilibrium.
The free evolution of the phonon subsystem is described by
\begin{subequations}
\begin{align}
H_{\t{Ph}}=\hbar\sum_\q \w_\q b_\q\+ b_\q\, ,
\end{align}
while the dot-phonon coupling reads\cite{Besombes2001,Machnikowski2004} 
\begin{align}
H_{\t{QD-Ph}}=\hbar\sum_\q
\left(\g_\q^{\t{X}}b_\q\+ +\g_\q^{\t{X}*}b_\q\right)\XX\, .
\end{align}
\end{subequations}
Here, the bosonic operator $b_\q\+$ ($b_\q$) creates (destroys) phonons with the frequency
$\w_\q$. $\g_\q^{\t{X}}$ denotes the deformation-potential-type coupling constant between
the exciton state and the $\q$-th bosonic mode.
Finally, the Hamiltonian for the photonic reservoir is given by
\begin{subequations}
\begin{align}
H_{\t{Rad}}=\hbar\sum_j \w_j d_j\+d_j\, ,
\end{align}
and the dot is radiatively coupled to the photons via
\begin{align}
  H_{\t{QD-Rad}}=\hbar\sum_j \left(\Gamma_j d_j\+ \GX +\Gamma_j^* d_j\XG\right)\, .
\end{align}
\end{subequations}
$d_j\+$ ($d_j$) creates (destroys) a photon with energy $\hbar\w_j$, where the coupling
constant is $\Gamma_j$.

Since the characteristic super-ohmic coupling of the phonons leads to pronounced
non-Markovian behavior\cite{Foerstner2003,Vagov2004,Thorwart2005,Vagov2007,Ramsay2010a,Kaer2010},
a simple Born-Markov approximation for the phononic environment is typically insufficient.
In contrast, the photonic memory is negligible compared to all time scales of the subsystem
of interest.
Therefore, it is a good approximation to treat the photonic reservoir as being Markovian.
The central result of this paper is to derive a practical algorithm to calculate
multi-time correlation functions using PI methods, which is capable of keeping
all memory effects induced by the non-Markovian environment (here represented by LA
phonons).
Apart from discretization errors the scheme does neither introduce approximations
with respect to the dot-laser interaction nor to the dot-phonon coupling.
Furthermore, we show how additional couplings to Markovian baths (in our example the
radiative coupling) can be accounted for consistently with the non-Markovian parts of the
dynamics by using a reduced description that eliminates the degrees of freedom corresponding
to the Markovian baths.

It is important to note that even when the Markovian interactions 
are included all memory effects related to the non-Markovian environment are still
retained.
Furthermore, even though the photon and the phonon coupling are evaluated in the bare
state basis of the QD, mutual photonic and phononic renormalizations are nevertheless
fully contained in the model.
These renormalizations build up during the propagation
since dot-phonon and dot-photon interactions do not commute, a property 
which still holds also for the Lindblad operators that are derived from the 
dot-photon coupling Hamiltonian.

\section{Evaluation of multi-time correlation functions}

A general multi-time correlation function can be defined as
\begin{align}
\label{eq:def_general_multi-time}
&G_{O_1...O_{2N}}(t_1,...,t_N)=\nn
&\<O_1(t_1)O_2(t_2)...O_N(t_N)O_{N+1}(t_N)...O_{2N-1}(t_2)O_{2N}(t_1)\>\, ,
\end{align}
where $O_1...O_{2N}$ are Hilbert space operators of the subsystem of interest (here the QD system)
in the Heisenberg picture and the times are ordered as $t_1\leq t_2\leq...\leq t_N$.
Note that this definition covers the common case where operator pairs with the same time argument
appear as, e.g., in the second-order two-time polarization correlation function
\begin{align}
G^{(2)}(t,\tau):=&\,G_{\s^{+}\s^{+}\s^{-}\s^{-}}(t,t+\tau)\nn
=&\,\<\s^{+}(t)\s^{+}(t+\tau)\s^{-}(t+\tau)\s^{-}(t)\>
\end{align}
with the polarization operators $\s^{+}=\XG$ and $\s^{-}=\GX$.
Also multi-time functions are covered by this definition,
where a given time argument is assigned to only one operator, as applies for the first-order
two-time correlation function $G^{(1)}(t,\tau):=\<\s^{+}(t+\tau)\s^{-}(t)\>$.
The latter can be obtained from the general expression Eq.~(\ref{eq:def_general_multi-time}) for $N=2$ by
setting two of the four appearing operators to the identity operator.
Note also that by definition the time dependence of the operators $O_{j}(t_{\ell})$ refers to the
Heisenberg picture and thus presumes the existence of a unitary time evolution operator.

In the following we consider $N=2$ to keep the notation concise and since
the major step from one- to multi-time functions is already captured in that
case. The generalization to arbitrary $N$ is straightforward and will be explained
at the end of the section.

\subsection{Tracing out the Markovian bath}
\label{sec:Makovian_bath}

The first step in order to obtain a reduced description of multi-time correlation functions is to
eliminate the degrees of freedom corresponding to the Markovian bath (here the photonic bath).
This reduction is most simple when single-time expectation values of operators not acting on the
photonic reservoir are considered, since all such expectation values can be evaluated
using the reduced density matrix
\begin{align}
\label{general_rdm}
\barr(t)=\Tr_{\t{Rad}}[\r(t)]\, ,
\end{align}
where $\Tr_{\t{Rad}}[\cdot]$ denotes the partial trace over the photonic subspace
giving rise to radiative decay.
Consider now the following four requirements:
(i) the continuum described by $H_{\t{Rad}}+H_{\t{QD-Rad}}$ can be treated within the
Markov approximation, (ii) system and environment subspaces factorize initially, i.e.,
$\r(0)=\barr(0)\otimes \r_{\t{Rad}}$, where $\r_{\t{Rad}}$ is a thermal distribution,
(iii) $H_{1}(t)$ does not depend on $d_j\+$ and $d_j$, and (iv) the frequency renormalization (Lamb shift)
induced by the mode continuum can be neglected as well as the finite temperature of the Markovian bath.
As discussed in textbooks \cite{Louisell1973,Carmichael1993}, if these requirements are fulfilled, the reduced density matrix obeys the dynamical equation
\begin{align}
\label{eq:rho_Liouville}
\frac{\partial}{\partial t}\barr(t)
=&\,\mathcal{L}\barr(t)\\
\t{with}\quad \mathcal{L}=&\,\mathcal{L}_{1}+\mathcal{L}_{\s^{-},\g}\, ,
\end{align} 
where $\mathcal{L}_{1}$ is defined by
\begin{align}
\mathcal{L}_{1}\barr=-\frac{i}{\hbar}[H_{1}(t),\barr]
\end{align}
and describes the Hamiltonian dynamics induced by $H_{1}(t)$.
The second term accounts for the interaction with the Markovian bath by the Lindblad superoperator $\mathcal{L}_{\s^{-},\g}$ which is constructed following the well known scheme for Lindblad operators:
\begin{align}
\mathcal{L}_{O,\g}\r=\frac{\g}{2}(2O\r O\+ -O\+ O\r-\r O\+ O)
\end{align}
with a loss rate $\g$ and a system operator $O$.
Provided that $\mathcal{L}_{\s^{-},\g}$ is non-zero, Eq.~(\ref{eq:rho_Liouville}) represents a non-unitary
time evolution in Liouville space.
The simultaneous treatment of such non-Hamiltonian and Hamiltonian dynamics within a PI
approach for single-time functions is discussed in Ref.~\onlinecite{Barth2016}, where the phonon 
subspace is traced out keeping its full memory structure.
Note that due to condition (iii) the phononic subspace is not affected by tracing out the Markovian bath,
i.e., it is still formulated as a Hamiltonian contribution to the dynamics.
Introducing the Liouvillian propagator
\begin{align}
\label{eq:Liouvillian_Propagator}
\mathcal{P}_{t'\to
t}=\mathcal{T}e^{\int_{t'}^{t}\mathcal{L}(t'')dt''}
\end{align}
with the time ordering operator $\mathcal{T}$, the formal solution of Eq.~(\ref{eq:rho_Liouville}) is
$\barr(t)=\mathcal{P}_{0\to t}\barr(0)$.

For the evaluation of two-time functions we start with the definition in Eq.~(\ref{eq:def_general_multi-time})
for the case $N=2$ where the time-evolution of the operators follows from the total Hamiltonian $H(t)$. 
Using the cyclic property of the trace, which is taken over the entire space considered, we can rearrange this expression to yield
\begin{align}
\label{eq:reordered}
G_{O_1O_2O_3O_4}(t_1,t_2)=
&\Tr_{\t{QD,Ph,Rad}}\Big[O_2(t_2)O_{3}(t_2)\nn
&\times O_{4}(t_1)\r(0)O_1(t_1)\Big]\, ,
\end{align}
where the indices QD, Ph, Rad indicate the decompostion of the total trace
into parts referring to the QD, the phonon
and the photon degrees of freedom, respectively.
Changing the dynamical picture with the help of the unitary time evolution operator 
$\mathcal{U}_{t'\to t}=\mathcal{T}e^{-\frac{i}{\hbar}\int_{t'}^{t}H(t'')dt''}$ leads to
\begin{align}
\label{eq:def_two-time}
&G_{O_1O_2O_3O_4}(t_1,t_2)=\nn
&\Tr_{\t{QD,Ph}}\Big[O_2(0)O_3(0)
\Tr_{\t{Rad}}\Big[\mathcal{U}_{t_{1}\to t_{2}}O_{4}(0)\r(t_1)
O_1(0)\mathcal{U}_{t_{1}\to t_{2}}\+\Big]\Big]\, .
\end{align}
Eq.~(\ref{eq:def_two-time}) is the starting point for the standard elimination of the Markovian bath,
where the central assumption is made that the total density matrix $\r(t_1)$ factorizes not only at
some initial time $t_0$ but at every single point in time $t_1$ into system and reservoir parts. 
Furthermore, the reservoir is taken to be static, i.e., it is described by the thermal distribution
$\r_{\t{Rad}}$ for all times.
When these conditions are fulfilled to a good approximation, the two-time correlation function can be written
as\cite{Louisell1973,Carmichael1993}
\begin{align}
\label{eq:QRT}
&G_{O_1O_2O_3O_4}(t_1,t_2)=\nn
=&\Tr_{\t{QD,Ph}}\Big[O_2(0)O_3(0)\mathcal{P}_{t_1\to t_2}
\Big[O_4(0)[\mathcal{P}_{0\to t_1}\barr(0)]O_1(0)\Big]\Big]\, ,
\end{align}
where $\mathcal{P}_{t'\to t}$ is the Liouville space propagator that has been introduced in Eq.~(\ref{eq:Liouvillian_Propagator}).
The advantage of using Eq.~(\ref{eq:QRT}) instead of Eq.~(\ref{eq:def_two-time}) is that
$\mathcal{P}_{t'\to t}$ acts in the reduced space, where the operators of the Markovian bath have been eliminated, while the unitary time evolution operators $\mathcal{U}_{t'\to t}$ act in the
unreduced total space.

We stress that, up to this point, the phononic environment is treated completely Hamiltonian.
In absence of a non-Markovian environment, Eq.~(\ref{eq:QRT}) would constitute the desired reduced description,
which in that case reduces to the QRT.

\subsection{Path integral treatment of the non-Markovian reservoir}
\label{sec:PI}

Eq.~(\ref{eq:QRT}) is a formal representation of the sought two-time correlation function.
The main target of the present paper is to provide a practical algorithm that enables a complete numerical
evaluation of $G_{O_1O_2O_3O_4}(t_1,t_2)$. For single-time functions this goal has been
reached before with the use of PI methods.\cite{Makri1995a,Makri1995b,Thorwart2000,Vagov2011,Barth2016}
The extension to two-time functions is most easily understood after first recalling how the
PI approach works for single-time functions.
The task to evaluate single-time expectation values of system operators $O_{j}$ (QD operators in our example)
can be split in first obtaining the reduced density matrix $\dbarr$ for the QD subsystem and then taking
the trace over $O_{j}\dbarr$.
As usual, the reduced density matrix of the QD subsystem is defined as:
\begin{align}
\dbarr(t)=\Tr_{\t{Ph}}\left[\barr(t)\right]\, ,
\end{align}
where the trace is taken over the phonon degrees of freedom.
In fact, the evaluation of the matrix elements $\dbarr_{\n\m}$ is a special case of Eq.~(\ref{eq:QRT}),
where $t_{1}=t_{2}=t$, $O_{2}=O_{3}=O_{4}=\t{Id}$ with $\t{Id}$ being the identity operator and
$O_{1}=|\m\>\<\n|$, such that
\begin{align}
\label{eq:one-time}
\dbarr_{\n\m}(t)=\Tr_{\t{QD,Ph}}\Big[[\mathcal{P}_{0\to t}\barr(0)]|\m\>\<\n|\Big]\, .
\end{align}
The expression in Eq.~(\ref{eq:one-time}) can be cast into a PI representation by decomposing the
Liouvillian propagator according to
\begin{align}
\label{eq:decomposition}
\mathcal{P}_{0\to t}=&\t{Id}^{(n)}\mathcal{P}_{(n-1)\D t\to n\D t}\t{Id}^{(n-1)}\cdots\nn
&\cdots\t{Id}^{(2)}\mathcal{P}_{\D t\to 2\D t}\t{Id}^{(1)}\mathcal{P}_{0\to \D t}\t{Id}^{(0)}\, .
\end{align}
Here we have subdivided the time interval $[0,t]$ into $n$ equidistant time steps $\D t$, such that $n\D t=t$.
Furthermore, the operators $\t{Id}^{(j)}$ that are inserted after the $j$-th time step are identity operators
that according to the completeness relation may be written as
\begin{align}
\label{eq:completeness}
\t{Id}^{(j)}=\sum_{\x_{j}}|\x_{j}\>\<\x_{j}| \, ,
\end{align}
where $\{|\x_{j}\>\}$ is a complete set of states in the QD and phonon subspace.
Inserting the identity operators in this form replaces the operators $\mathcal{P}_{(j-1)\D t\to j\D t}$ by their
matrix elements and thus the reduced density matrix in Eq.~(\ref{eq:one-time}) appears as a sum over paths in
configuration space of QD and phonons.
It is well known that all parts of these summations involving phonons can be performed analytically for the
pure-dephasing coupling model\cite{Makri1995a,Makri1995b,Barth2016}, such that in the numerical treatment only
summations over paths in the QD configuration space have to be carried out.
Note that the classification of the coupling as being of pure-dephasing type refers to the bare state basis of the QD, where this coupling is diagonal.
Working in a QD basis, where the coupling to phonons is diagonal, simplifies greatly the task of analytically performing the required integrations over the phonon degrees of freedom, which eventually leads to the phonon memory kernels.
However, even for a two-level system
the direct summation over all paths is feasible only for few time steps, since the number of contributions grows
exponentially with the number of time steps.

A highly efficient algorithm for performing the sum over paths has been introduced by Makri and
Makarov\cite{Makri1995a,Makri1995b}, which makes use of the fact that after integrating out the phonon degrees
of freedom, their influence is captured by phonon induced memory kernels that for a continuum of phonons decay
on a finite time scale (for LA phonon coupling to a QD typically on the order of a few picoseconds).
Note that the phonon induced memory kernels can be evaluated from the knowledge of the phonon coupling 
constants $\g_{\q}^{X}$ and thus the memory depth is known before starting any path summations. Using the
finite memory depth the summation over paths can be carried out by an iterative scheme for the so called
augmented density matrix (ADM).
The ADM is a $2n_{c}$ dimensional tensor of weights assigned to each path where $n_{c}$ is the number of
time steps needed to cover the full memory depth. In the case where the Liouville propagator contains
simultaneously Hamiltonian and non-Hamiltonian contributions, the calculation of the ADM proceeds
according to the following recursion\cite{Barth2016}:
\begin{align}
\label{eq:recursion}
&\dbarr_{\n_n...\n_{n-n_c+1}}^{\m_n...\m_{n-n_c+1}}
=\mathcal{M}_{\n_n\m_n}^{\n_{n-1}\m_{n-1}}\nn
&\times
\sum_{\substack{\n_{n-n_c}\\\m_{n-n_c}}}
\exp{\left(\sum_{l=n-n_{c}}^{n}S_{\n_{n}\n_{l}\m_{n}\m_{l}}\right)}
\dbarr_{\n_{n-1}...\n_{n-n_c}}^{\m_{n-1}...\m_{n-n_c}}\, ,
\end{align}
where $S_{\n_{n}\n_{l}\m_{n}\m_{l}}$ is the phonon influence functional that describes the time-delayed
back action of the phononic environment, in short, the phonon-induced memory of the system.
$\mathcal{M}_{\n_n\m_n}^{\n_{n-1}\m_{n-1}}$ is the matrix element of the Liouville propagator
$\mathcal{M}_{t}$ representing the combined time evolution induced in the QD subsystem\cite{Barth2016,Cygorek2017} 
by $H_{0}(t)$ and $\mathcal{L}_{\s^{-},\g}$ from time step $n-1$ to $n$.
For the sake of a self-contained presentation, we repeat the explicitly known expressions for
$S_{\n_{n}\n_{l}\m_{n}\m_{l}}$ and $\mathcal{M}_{\n_n\m_n}^{\n_{n-1}\m_{n-1}}$ in Appendix~\ref{sec:app}.
The iteration is started by calculating the ADM for the first $n_{c}$ time steps
by directly performing the corresponding summations.\cite{Vagov2011,Cygorek2017}

To obtain the reduced density matrix $\dbarr_{\n\m}$ from the ADM one has to trace out all indices
except those referring to the most recent time step, i.e.,
\begin{align}
\dbarr_{\n_n\m_n}
=&\sum_{\substack{\n_{n-1}...\n_{n-n_c+1}\\\\\m_{n-1}...\m_{n-n_c+1}}}
\dbarr_{\n_n...\n_{n-n_c+1}}^{\m_n...\m_{n-n_c+1}}\, .
\end{align}
In order to develop a PI-based algorithm for the evaluation of the two-time function in Eq.~\eqref{eq:QRT}
we proceed along the same lines as done for the evaluation of the single-time reduced density matrix,
i.e., in a first step we decompose in Eq.~\eqref{eq:QRT} both $\mathcal{P}_{0\to t_{1}}$ as well as $\mathcal{P}_{t_{1}\to t_{2}}$
according to Eq.~\eqref{eq:decomposition}.
As for the single-time functions the insertion of identity operators resolved by the completeness relation
Eq.~\eqref{eq:completeness} casts the expression Eq.~\eqref{eq:QRT} into a sum over paths in configuration space.
It should be noted that the operators $O_{j}$ in Eq.~\eqref{eq:QRT} act only within the QD subspace and
not on the phonon degrees of freedom.
Therefore, all integrals over phonon variables that show up after inserting the completeness relation
Eq.~\eqref{eq:completeness} for the identity operators are identical to those appearing in the
corresponding expressions for single-time functions.
In particular, all related integrations can be carried out analytically such that the phonon influence
again ends up in memory kernels that can be evaluated beforehand and thus the memory depth is known.
For efficiently performing the summations over paths an iterative scheme analogous to
Eq.~\eqref{eq:recursion} may be set up. To this end we assume a time discretization
such that $n$ equidistant time steps of length $\D t$ are needed to cover the interval
$[0,t_{1}]$ while another $m$ time steps of the same length connect $t_{1}$ with $t_{2}$, i.e.,
$t_{1}=n\D t$ and $\tau_{1}:=t_{2}-t_{1}=m\D t$.
From Eq.~\eqref{eq:QRT} it can be seen that in order to reach $t_{1}$ starting from $t=0$ the ADM
can be iterated in the same way as for single-time functions, i.e., according to Eq.~\eqref{eq:recursion}. 
Applying the operator $O_{4}$ from the left and $O_{1}$ from the right to the operator
$\mathcal{P}_{0\to t_1}\barr(0)$
is formally equivalent to replacing the operator $\mathcal{M}_{t}$, which describes the
time evolution of the QD subsystem without phonon coupling from $t_1$ to $t_1+\D t_1$,
in the $n$-th time step 
by $O_{4}(0)\mathcal{M}_{t_1}O_{1}(0)$.
We can therefore define a modified ADM which for the first $n$ time steps coincides with the original
ADM.
At the $n$-th time step the modified ADM is defined as:
\begin{align}
\label{eq:mod_ADM}
&\dbarr_{O_4O_1\n_n...\n_{n-n_c+1}}^{\qquad\m_n...\m_{n-n_c+1}}
=\sum_{\n'_n\m'_n}
(O_4)_{\n_n\n'_n}
\mathcal{M}_{\n'_n\m'_n}^{\n_{n-1}\m_{n-1}}
(O_1)_{\m'_n\m_n}\nn
&\times
\sum_{\substack{\n_{n-n_c}\\\m_{n-n_c}}}
\exp{\left(\sum_{l=n-n_{c}}^{n}S_{\n_{n}\n_{l}\m_{n}\m_{l}}\right)}
\dbarr_{\n_{n-1}...\n_{n-n_c}}^{\m_{n-1}...\m_{n-n_c}}\, ,
\end{align}
where $(O_1)_{\n\m}$ and $(O_4)_{\n\m}$ are the representations of the system operators in the QD basis
$\{|\n\>\}$ obtained by inserting the completeness relation with respect to this basis between
$O_1$ ($O_4$) and $\mathcal{M}_{t}$.
From Eq.~\eqref{eq:QRT} it is evident that after applying the operators $O_{1}$ and $O_{4}$ at time $t_{1}$
there is no further modification compared with the single-time case during the propagation from time $t_{1}$
to $t_{2}=t_{1}+\tau_{1}$ and thus the modified ADM $\dbarr_{O_4O_1}$ obeys the same recursion as the ADM,
i.e., Eq.~\eqref{eq:recursion}, for the subsequent $m$ time steps covering the interval $[t_{1},t_{2}]$.

Finally, the calculation of the two-time function $G_{O_1O_2O_3O_4}(t_1,t_2)$ is completed by
applying $O_{2}(0)O_{3}(0)$ from the left and then taking the trace over the QD degrees of freedom, i.e.,
\begin{align}
&G_{O_1O_2O_3O_4}(t_1,t_2) =\nn
& \sum_{\substack{\n_{n+m}...\n_{n+m-n_c+1}\\\\\m_{n+m}...\m_{n+m-n_c+1}}}\hspace*{-10mm}
\Big[O_{2}(0)O_{3}(0)\Big]_{\m_{n+m}\n_{n+m}}\dbarr_{O_4O_1\n_{n+m}...\n_{n+m-n_c+1}}^{\qquad\m_{n+m}...\m_{n+m-n_c+1}}\, ,
\end{align}
where $\Big[O_{2}(0)O_{3}(0)\Big]_{\m\n}$ is the matrix element of the operator product $O_{2}(0)O_{3}(0)$
in the QD basis.
It is important to note that for evaluating the two-time function $G_{O_1O_2O_3O_4}(t_1,t_2)$ for 
finite delays $\tau_{1}=t_{2}-t_{1}>0$ it is not sufficient to know the reduced density matrix
at time $t_{1}$.
This is the crucial difference compared with the QRT where the two-time function is
propagated with respect to the delay $\tau_{1}$ by solving an initial value problem where the initial values are
determined from the reduced density matrix at $t_{1}$. The physical meaning of this difference is that
iterating the modified ADM keeps the complete phonon induced memory before and after the time $t_{1}$.

Altogether, the above described iteration scheme for performing a PI-based calculation of the two-time function
$G_{O_1O_2O_3O_4}(t_1,t_2)$ requires $n+m$ iterations, the effort of which is the same as needed for the
iteration of the ADM in the single-time case.
Thus, the evaluation of $G_{O_1O_2O_3O_4}(t_1,t_2)$ for a given pair $t_{1}$, $t_{2}$ 
can be done with the same numerical effort as needed for tracing a single-time function over a time interval
$[0,t_{1}+\tau_{1}]$.
Of course, this implies that mapping out the dependence on both $t_{1}$ and $\tau_{1}$
is much more demanding than following a single-time function over an extended time interval since
the calculation has to be performed anew for each pair $t_{1},\tau_{1}$.

Finally, we would like to note that the generalization to an arbitrary multi-time function as defined in
Eq.~(\ref{eq:def_general_multi-time}) can easily be done by repeating the derivations of
subsections~\ref{sec:Makovian_bath} and \ref{sec:PI} for every time argument $t_j$ with $2<j\leq N$ for
the subsequent dynamics in $\tau_j$. This amounts to (i) reordering the operators in
Eq.~\eqref{eq:def_general_multi-time} such that operators with equal times appear on either side
of the density matrix as in Eq.~\eqref{eq:reordered}, (ii) then, if present, tracing out Markovian
baths which converts the unitary time evolution into a Liouville space propagation described by
operators $\mathcal{P}_{t_{j}\to t_{j+1}}$, (iii) decomposing these operators according to
Eq.~\eqref{eq:decomposition}, (iv) obtaining a representation of the multi-time
function as a sum over paths in configuration space by resolving the identity operators introduced
in the decomposition using the completeness relation Eq.~\eqref{eq:completeness},
(v) representing the phonon induced memory in terms of memory kernels by performing the
integrations over phonon degrees of freedom , (vi) performing
the remaining sum over paths in the QD subspace by iterating a modified ADM which at every intermediate
time argument $t_{j}$ of the $N$-time function
picks up operators $O_{j}$ and $O_{2N-j+1}$, (vii) finally, after the last time step applying operators
$O_{N}$ and $O_{N+1}$ to the modified ADM and taking the sum over all system indices of
the modified ADM.
Again for a given ordered tuple $(t_{1},\cdots,t_{N})$ the numerical demand is identical to
the effort of following a single-time function from time $t=0$ to the final time $t=t_{N}$.
This is due to the fact that - as a simple recipe - the derivation amounts to modifying
the Liouvillian propagator $\mathcal{M}_t$, that describes the combined Hamiltonian and
non-Hamiltonian evolution of the subsystem of interest, at times $t\in\{t_1,...,t_{N-1}\}$
in complete analogy to Eq.~\ref{eq:mod_ADM}.

\begin{figure*}
\begin{minipage}[t]{0.49\textwidth}
\begin{center}
\includegraphics[width=\textwidth]{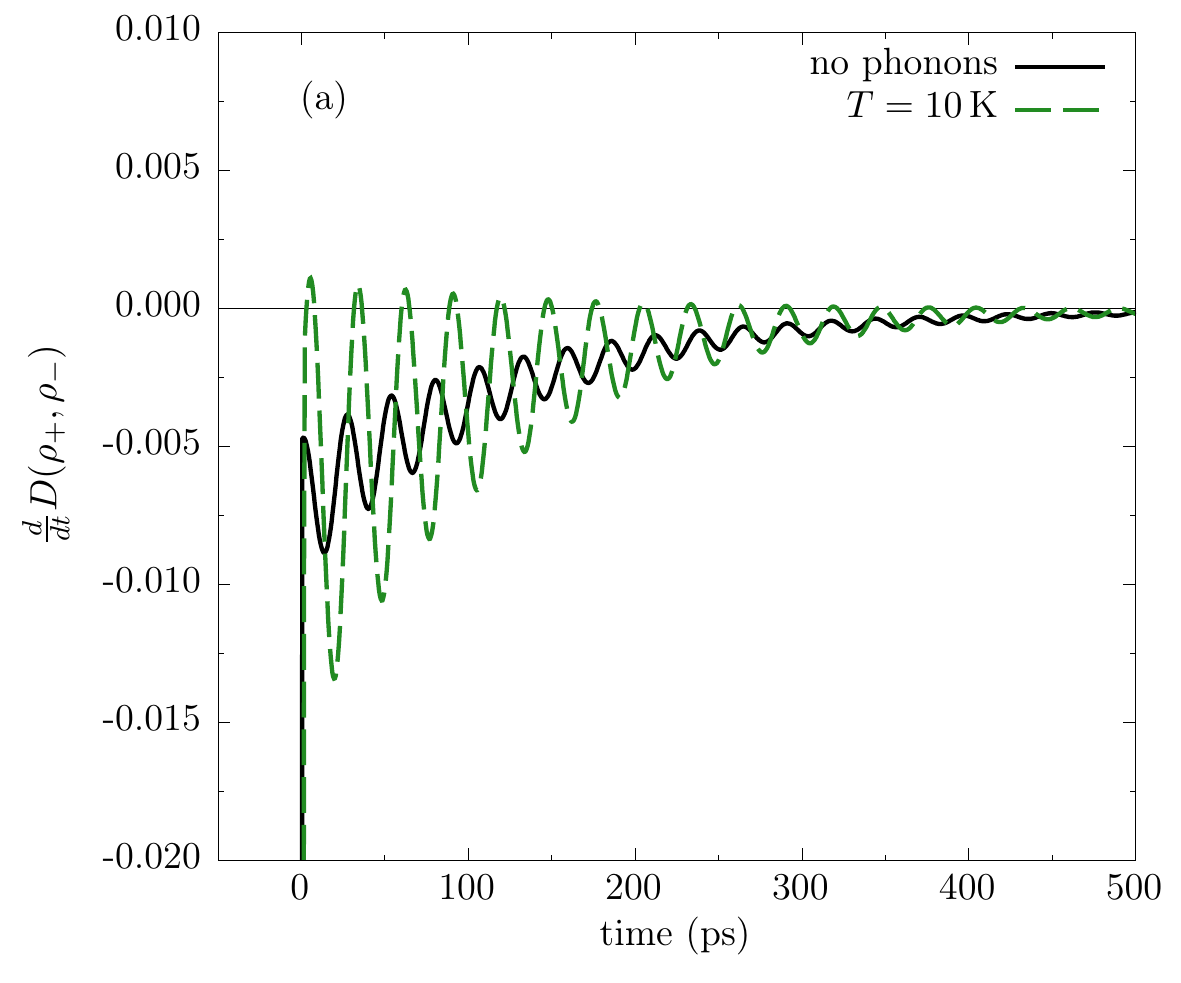}
\end{center}
\end{minipage}
\hfill
\begin{minipage}[t]{0.49\textwidth}
\begin{center}
\includegraphics[width=\textwidth]{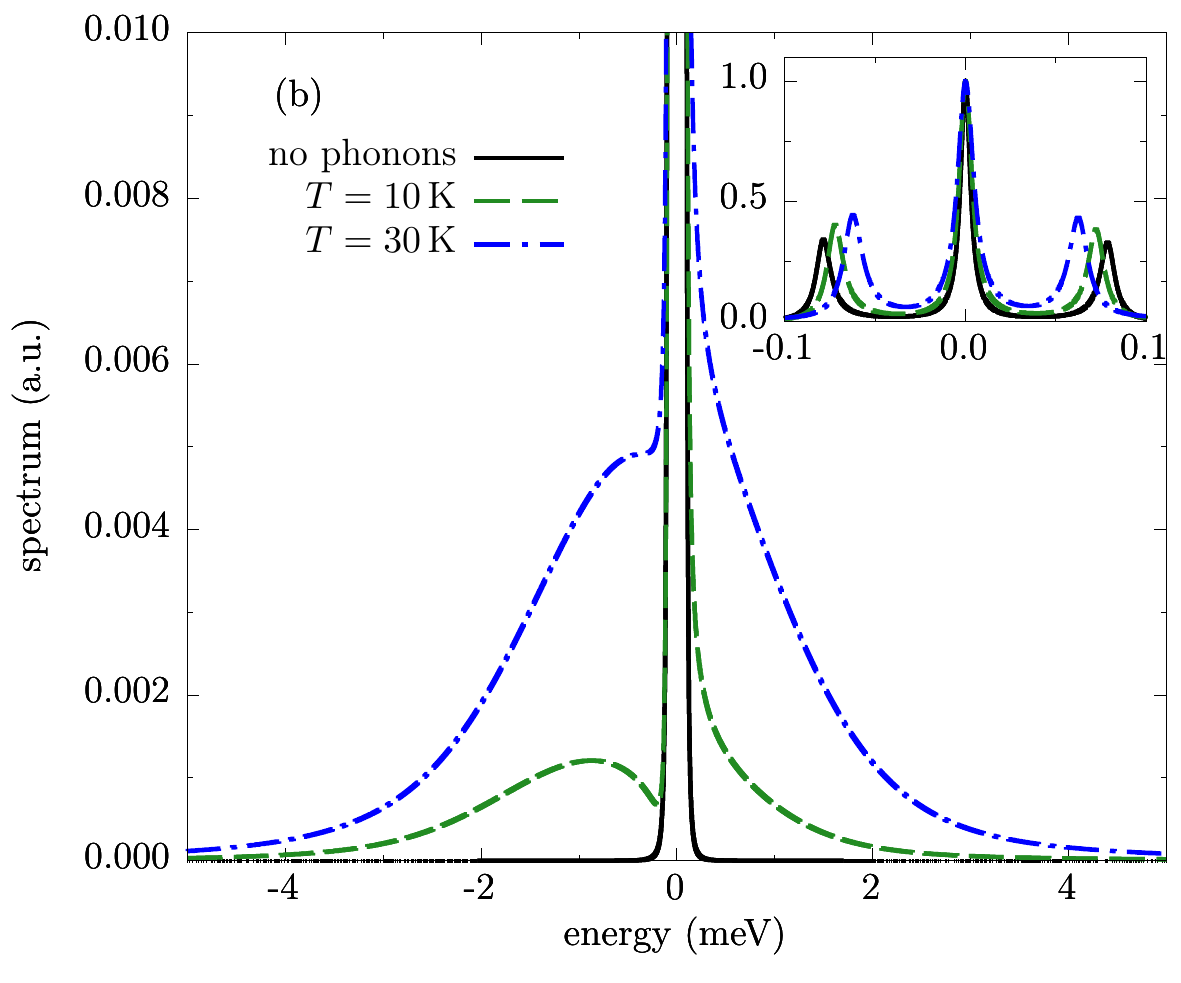}
\end{center}
\end{minipage}
\caption{(a) Derivative of the trace distance and (b) incoherent resonance fluorescence,
i.e., emission spectrum including phonons initially at $T=10\,\t{K}$, $T=30\,\t{K}$, and without phonons
for comparison for a $6\,\t{nm}$ GaAs QD with cw excitation. The inset shows the same data on a scale,
where the Mollow triplet is visible. The spectra are normalized to the height of the central peak.}
\label{fig:Trace_Distance_and_Spectrum}
\end{figure*}

\section{Numerical examples}
\label{sec:Numerical_Example}

In order to illustrate the applicability of our algorithm we first consider a
laser driven two-level system coupled to LA phonons and accounting for radiative
decay by a Lindblad operator derived from a Markovian photonic bath as described in
section~\ref{sec:Model}.
For the explicit calculation we take a GaAs-based self-assembled dot with a diameter of
$6\,\t{nm}$ and standard material parameters.\cite{Krummheuer2005}
The laser is tuned in resonance to the two-level transition and, after being switched on at $t=0$,
has a constant amplitude corresponding to $\hbar f(t)/2=39\,\t{\textmu eV}$.
The radiative decay rate is taken to be $\hbar\g=6.6\,\t{\textmu eV}$.
The phonons are assumed to be initially in thermal equilibrium at a temperature of $T=10\,\t{K}$.

Before we discuss two-time correlation functions of this system, we would like to demonstrate that
this system indeed exhibits pronounced non-Markovian dynamics for the chosen parameter set.
To this end, we show the time derivative of the trace distance $\frac{d}{dt}D(\r_{+},\r_{-})$,
as defined in Refs.~\onlinecite{Breuer2009,McCutcheon2016}, in
Fig.~\ref{fig:Trace_Distance_and_Spectrum}(a), which can be interpreted as a measure
of non-Markovianity.\cite{Breuer2009,McCutcheon2016}
According to Ref.~\onlinecite{Breuer2009} the Markovianity of a system implies that
$\frac{d}{dt}D(\r_{+},\r_{-})$ be less than zero.
In other words, a value larger than zero for this quantity indicates non-Markovian environment influences.
There is an intuitive interpretation to this criterion:
When the distance, i.e., the distinguishability between two states, increases with time
$\left(\frac{d}{dt}D(\r_{+},\r_{-})>0\right)$ there has to be a flow of information from the environment
to the subsystem of interest. Thus, the system dynamics cannot be Markovian.
For the application of this criterion it is sufficient to find a pair of states $\r_+$ and $\r_-$ for whom this statement holds.
Here, we choose the same states as defined in Ref.~\onlinecite{McCutcheon2016}.
Clearly, non-Markovian behavior is established in Fig.~\ref{fig:Trace_Distance_and_Spectrum}(a).
The origin of the non-Markovian dynamics is the phononic environment as can be seen by
switching off the dot-phonon coupling which results in values of $\frac{d}{dt}D(\r_{+},\r_{-})$
strictly lower than zero [cf. the black solid line in Fig.~\ref{fig:Trace_Distance_and_Spectrum}(a)].

\subsection{Emission spectrum of a two-level QD}

Next we turn to the calculation of the emission spectrum $S(\w)$ of our driven two-level system\cite{Reiter2017}
which we evaluate from the two-time correlation function 
$G^{(1)}(t,\tau)=\<\s^{+}(t+\tau)\s^{-}(t)\>$
by considering times $t$ long enough such that a stationary nonequilibrium state has been reached.
Then the Fourier transform is taken with respect to $\tau$ after subtracting the limiting value 
$\lim_{t\to\infty}\lim_{\tau\to\infty}G^{(1)}(t,\tau)$
which in Ref.~\onlinecite{Carmichael1993} is referred to as {\em coherent part} of the emission, i.e.,
\begin{align}
&S(\w)=\lim_{t\to\infty}\t{Re}\Big[\int_{-\infty}^{\infty}d\tau\,\Big(G^{(1)}(t,\tau)\nn
&-\lim_{\tau\to\infty}G^{(1)}(t,\tau)\Big)e^{-i\w\tau}\Big]\, .
\end{align}
The determination of the emission spectrum can be considered to be a crucial test of our method
since in Ref.~\onlinecite{McCutcheon2016} it has been demonstrated that a naive application of
the QRT yields the phonon side band energetically on the wrong side.
Our result is shown in Fig.~\ref{fig:Trace_Distance_and_Spectrum}(b), where spectra calculated
for phonons initially at $T=10\,\t{K}$ and $T=30\,\t{K}$ as well as for the phonon-free
case are shown.
Clearly, a broad phonon side band is found below the zero phonon line (ZPL) corresponding to
transitions involving phonon emission, whereas above the ZPL a tail is found originating from
transitions where phonons are absorbed. \cite{Iles-Smith2017, Iles-Smith2017b}
As expected, the tail is reduced with decreasing temperature while the broad phonon side band remains.
Furthermore, the same data plotted on a different scale [cf. inset in
Fig.~\ref{fig:Trace_Distance_and_Spectrum}(b)] shows the well known Mollow triplet.\cite{Mollow1969}
The Rabi splitting is due to the strong driving of $\hbar f=78\,\t{\textmu eV}$ compared with the
radiative decay rate. Also clearly visible is the phonon-induced renormalization and damping
of the Rabi oscillations that manifest themselves in the shift of the two side peaks towards the central
peak of the Mollow triplet and their broadening, respectively.
\cite{Foerstner2003,Machnikowski2004,Kruegel2005,Vagov2007,Nazir2008,Ramsay2010b,Wei2014}
Altogether, this demonstrates that our method reliably accounts for the phonon induced memory which
is reflected by the phonon-induced features in the spectra that appear on the energetically
correct side of the ZPL.

\begin{figure}
\begin{center}
\includegraphics[width=0.5\textwidth]{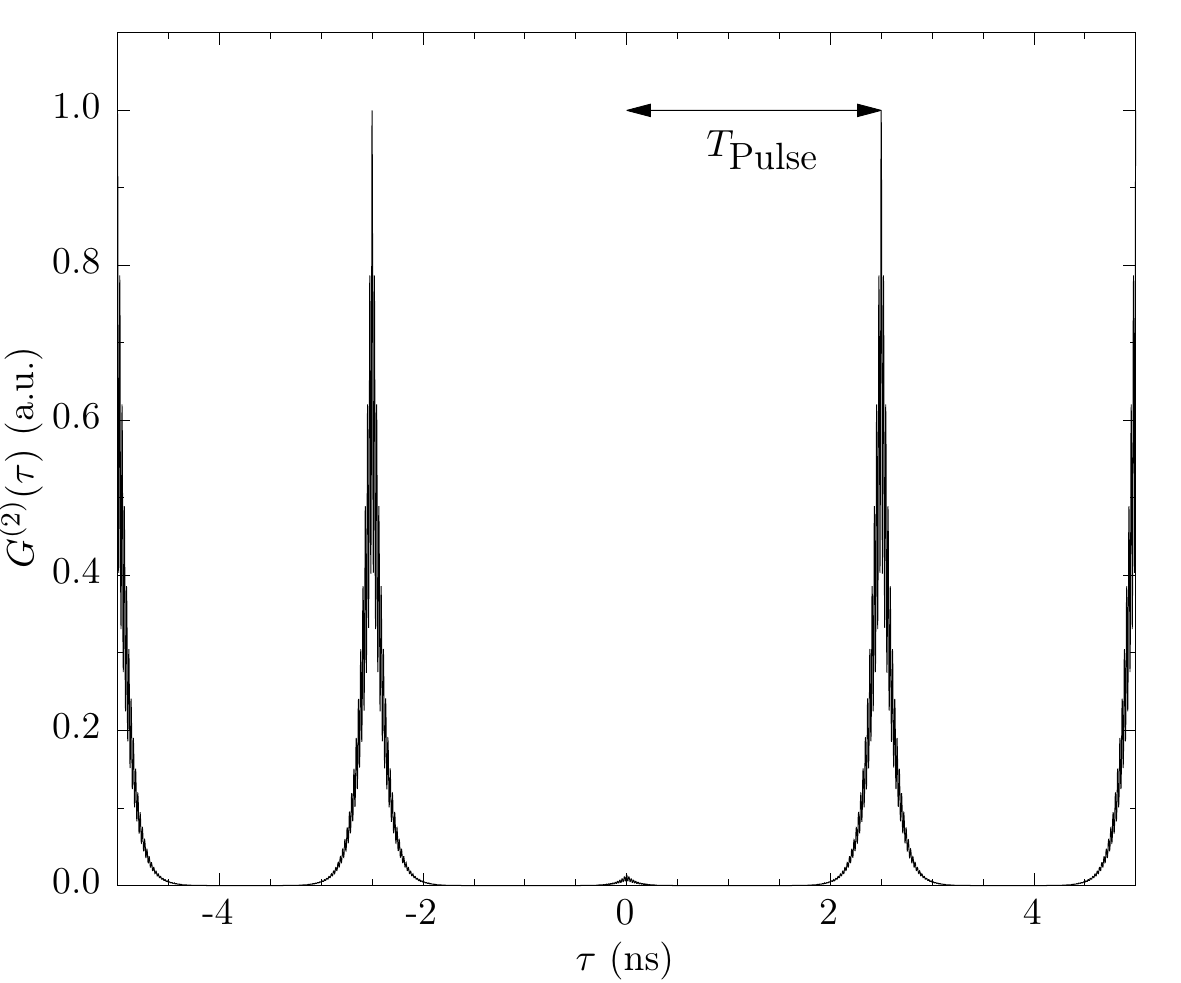}
\caption{Averaged two-time photon correlation function at $T=4.2\,\t{K}$ for a $6\,\t{nm}$ GaAs
QD embedded in a single-mode microcavity with pulsed excitation accounting for seven states on the
Jaynes-Cummings ladder.}
\label{fig:G2}
\end{center}
\end{figure}

\subsection{Two-time photonic correlation function of a QD-cavity system}

Usually two-level systems are used to showcase the application of PI methods 
because the rapid increase of the number of elements of the ADM with rising
number of levels of the system of interest imposes severe limitations to the size of 
systems that can be dealt with numerically. 
For QD systems, which exhibit typically a rather long memory of the order of a few picoseconds,
already for single-time functions corresponding calculations have been mostly restricted to
two-, three- or four-level systems.
\cite{Thorwart2005,Vagov2007,Glaessl2012,Barth2016b,Nahri2017}
Recently, it has been demonstrated in Ref.~\onlinecite{Cygorek2017} that for an important class
of systems, where the levels of the system of interest can be subdivided into groups with
identical phonon coupling within each group, the iterative summation over paths can be
speeded up enormously by iterating instead of the full ADM a quantity where the partial sum
over states within each group has already been performed.
Indeed, in Ref.~\onlinecite{Cygorek2017} a QD-cavity system with 41 states of the
Jaynes-Cummings ladder  coupled to LA phonons and accounting for cavity losses has
been treated.
This has become possible because in this case the reformulation reduced the numerical effort by
more than 15 orders of magnitude without introducing approximations.
It is easily seen that the idea of iterating a partially summed ADM can be combined 
with our algorithm for obtaining multi-time correlation functions without conflict.

To show the feasibility of such an enhanced scheme which paves the way toward numerically complete
calculations of multi-time correlation functions for systems beyond the very few level limit,
we have considered a laser driven two-level QD with a Jaynes-Cummings type coupling to a cavity mode 
with photon rising (lowering) operator $a\+$ ($a$)
and deformation potential coupling to LA phonons.
This amounts to considering an additional Hamiltonian contribution to the previously defined $H_0$ in the form of
\begin{align}
\label{eq:H_Cav}
H_{\t{Cav}}=\hbar\D\w_{CL}a\+ a + \hbar g\left(a\+ \GX + a \XG\right)\, ,
\end{align}
where $\D\w_{\t{CL}}$ is the detuning between the cavity mode and the central laser frequency.
In addition we account for cavity losses by a Lindblad term $\mathcal{L}_{a,\k}$ with a cavity
loss rate $\k$.
Our system of interest consists of the states of the Jaynes-Cummings ladder.
We show results where the system is driven by a pulse train $f(t)$ where every
$T_{\t{Pulse}}=2.5\,\t{ns}$ a laser pulse with Gaussian envelope $e(t)=\Theta/(\sqrt{2\pi}\sigma)\exp(-t^2/(2\sigma^2))$ hits the QD, i.e.,
\begin{align}
f(t)=\sum_{n=0}^\infty e(t+nT_{\t{Pulse}})\, .
\end{align}
The pulse area $\Theta$ is set to $\pi$ and the full
width at half maximum to $1\,\t{ps}$. 
The laser is in resonance with the phonon-shifted two-level transition as well as the
cavity mode.
The dot-cavity coupling is set to $\hbar g=100\,\t{\textmu eV}$ and the cavity loss rate is
chosen to be $\hbar\k=6.6\,\t{\textmu eV}$ while the temperature is taken to be
$T=4.2\,\t{K}$.
The radiative decay rate is $\hbar\g=6.6\,\t{\textmu eV}$ as before.
For these parameters converged results are obtained by keeping states with up to three excitations on
the Jaynes-Cummings ladder, which is a system with seven levels, namely three pairs of states $|G,n_{\t{X}}\>$, $|X,n_{\t{X}}-1\>$ for excitation numbers $n_{\t{X}}$ in the range $1\leq n_{\t{X}}\leq3$ and one state $|G,0\>$ for $n_{\t{X}}=0$.
Here, $|G,n\>$ and $|X,n\>$ denote product states where the dot is in one of its two energy eigenstates ($\G$ or $\X$) and $n$ photons are excited.
The excitation number $n_{\t{X}}$ refers to the sum of excited photons and excitons.
In the Jaynes-Cummings model (i.e., a system without phonons, cavity or radiative losses and without laser driving) $n_{\t{X}}$ is preserved.

The pulse driven QD-cavity systems under discussion are of particular interest concerning applications
as single-photon emitters.
It is possible to determine the quality of the emitted photons with
respect to this goal by examining the nonclassical antibunching behavior of the photonic subsystem.
To this end, the second-order two-time correlation function of the cavity photons is often considered
both in experiment and in theoretical modeling.
It is defined as
\begin{align}
\label{eq:def_G^2}
G^{(2)}(t,\tau):=\<a\+(t)a\+(t+\tau)a(t+\tau)a(t)\>\, ,
\end{align}
i.e., we set $N=2$ in Eq.~(\ref{eq:def_general_multi-time}) as well as $O_1=O_2=a\+$, $O_3=O_4=a$,
$t_1=t$, and $t_2=t+\tau$. 
$G^{(2)}(t,\tau)$ averaged over the first time argument is shown in Fig.~\ref{fig:G2}.
As expected, the dot-cavity system displays pronounced antibunching behavior.
The small amplitude modulations on top of the pronounced peak structure occur because
we are in the strong coupling limit where the QD-cavity coupling induces Rabi-type
oscillations.
Overall, our results confirm the feasibility of reliable numerically complete PI
calculations for systems noticeably larger than two-level systems.

\section{Conclusion}

We have presented a path-integral based algorithm for evaluating nonequilibrium 
multi-time correlation functions of a few-level subsystem that has a pure-dephasing type coupling
to a continuum of oscillators which may induce non-Markovian dynamics.
The algorithm is formulated in a way that allows to account for non-Hamiltonian contributions to the dynamics resulting from further couplings to Markovian baths in addition to the
Hamiltonian coupling to the oscillators.
In contrast to calculations of multi-time correlation functions based on the quantum
regression theorem, which is valid only for systems with Markovian dynamics, our approach keeps
the full memory structure induced by the oscillator continuum.
The algorithm can be applied to situations with explicit time dependence, e.g., due to external
laser driving, as well as to initial value problems.
Apart from limitations arising from the assumptions
made in the formulation of the model and, if applicable, apart from
making use of a Markovian description for some of the bath interactions,
no approximations are made in our treatment except for discretization errors.
In particular, all interactions are treated completely and non-perturbatively.
For example, for a quantum dot coupled to phonons all multi-phonon processes are included as
well as the laser driving to all orders in the driving field.
We have demonstrated that our method passes the crucial test case of determining the
emission spectrum of a two-level quantum dot coupled to LA phonons by showing that the
phonon side band appears on the correct side of the zero phonon line in contrast to results
obtained from the naive application of
the quantum regression theorem. 
In addition we have shown that the algorithm can be combined with a recent improvement of the
iteration scheme for performing the sum over paths which opens up new opportunities by
making numerically complete calculations of multi-time correlation functions feasible 
for systems where the number of relevant levels is considerably larger than two.
As an example for the latter situation we have shown converged results for the second-order
photonic correlation function evaluated for a seven-level system.

\acknowledgments
M. Cygorek thanks the Alexander-von-Humboldt foundation for support through a Feodor Lynen fellowship.
A. Vagov acknowledges the support from the Russian Science Foundation 
under the Project 18-12-00429.

\appendix
\section{Explicit expressions for the influence functional and Liouvillian propagator matrices}
\label{sec:app}

The following definitions are a repetition of those in Ref.~\onlinecite{Barth2016} for the sake
of a self-contained presentation of the present paper.
In the special case where the dot-phonon coupling constants are all either purely real or purely
imaginary (which applies to the deformation potential coupling to LA phonons),
the phonon influence fuctional reads
\begin{align}
S_{\n_l\n_{l'}\m_l\m_{l'}}=&\,-K_{\n_{l'}\n_l}(t_l-t_{l'})-K_{\m_{l}\m_{l'}}^*(t_l-t_{l'})\nn
&\,+K_{\n_{l}\m_{l'}}^*(t_l-t_{l'})+K_{\n_{l'}\m_l}(t_l-t_{l'})
\end{align}
with the memory kernels
\begin{align}
K_{\n_l\m_{l'}}(\tau)=&\,2\int_0^\infty\,d\w\frac{J_{\n_l\m_{l'}}(\w)}{\w^2}(1-\cos{(\w\D t))}\nn
&\,\times\left[\coth{\left(\frac{\hbar\w}{2k_BT}\right)}\cos{(\w \tau)}-i\sin{(\w\tau)}\right]
\end{align}
and
\begin{align}
K_{\n_l\m_{l}}(0)=&\,\int_0^\infty\,d\w\frac{J_{\n_l\m_{l}}(\w)}{\w^2}\nn
&\,\times\Big[\coth{\left(\frac{\hbar\w}{2k_BT}\right)}(1-\cos{(\w \D t)})\nn
&\,+i\sin{(\w\D t)}-i\w\D t\Big]\, .
\end{align}
Here, the phonon spectral density
\begin{align}
J_{\n\m}(\w)=\sum_\q\g_\q^\n\g_\q^{\m*}\delta(\w-\w_\q)
\end{align}
has been introduced.
Although in our case the spectral density is of the super-Ohmic type (cf. Ref.~\onlinecite{Vagov2011}), we would like to mention that various spectral densities describing vastly different systems may lead to pronounced non-Markovian behavior, e.g., nonanalyticities of the spectral function can induce power law tails in multi-time observables.\cite{Chakraborty2018}

Defining the combined Hamiltonian and non-Hamiltonian Liouvillian
for the subsystem of interest
\begin{align}
\mathcal{L}_{\t{S}}\bullet=-\frac{i}{\hbar}\left[H_0(t),\bullet\right]+\mathcal{L}_{\s^{-},\g}\bullet
\end{align}
the propagator
\begin{align}
\mathcal{M}_t\bullet=\mathcal{T}\exp{\left(\int_t^{t+\D t}\mathcal{L}_{\t{S}}\,dt'\right)}\bullet
\end{align}
is introduced. The matrix elements
\begin{align}
\mathcal{M}_{\n_l\m_l}^{\n_{l-1}\m_{l-1}}=\<\n_l|\mathcal{M}_t\Big[|\n_{l-1}\>\<\m_{l-1}|\Big]|\m_l\>
\end{align}
are defined as the elements of the operator that is obtained by applying
the propagator $\mathcal{M}_t$ to the canonical basis operator $|\n_{l-1}\>\<\m_{l-1}|$.
\bibliography{bib}

\begin{thebibliography}{60}%
\makeatletter
\providecommand \@ifxundefined [1]{%
 \@ifx{#1\undefined}
}%
\providecommand \@ifnum [1]{%
 \ifnum #1\expandafter \@firstoftwo
 \else \expandafter \@secondoftwo
 \fi
}%
\providecommand \@ifx [1]{%
 \ifx #1\expandafter \@firstoftwo
 \else \expandafter \@secondoftwo
 \fi
}%
\providecommand \natexlab [1]{#1}%
\providecommand \enquote  [1]{``#1''}%
\providecommand \bibnamefont  [1]{#1}%
\providecommand \bibfnamefont [1]{#1}%
\providecommand \citenamefont [1]{#1}%
\providecommand \href@noop [0]{\@secondoftwo}%
\providecommand \href [0]{\begingroup \@sanitize@url \@href}%
\providecommand \@href[1]{\@@startlink{#1}\@@href}%
\providecommand \@@href[1]{\endgroup#1\@@endlink}%
\providecommand \@sanitize@url [0]{\catcode `\\12\catcode `\$12\catcode
  `\&12\catcode `\#12\catcode `\^12\catcode `\_12\catcode `\%12\relax}%
\providecommand \@@startlink[1]{}%
\providecommand \@@endlink[0]{}%
\providecommand \url  [0]{\begingroup\@sanitize@url \@url }%
\providecommand \@url [1]{\endgroup\@href {#1}{\urlprefix }}%
\providecommand \urlprefix  [0]{URL }%
\providecommand \Eprint [0]{\href }%
\providecommand \doibase [0]{http://dx.doi.org/}%
\providecommand \selectlanguage [0]{\@gobble}%
\providecommand \bibinfo  [0]{\@secondoftwo}%
\providecommand \bibfield  [0]{\@secondoftwo}%
\providecommand \translation [1]{[#1]}%
\providecommand \BibitemOpen [0]{}%
\providecommand \bibitemStop [0]{}%
\providecommand \bibitemNoStop [0]{.\EOS\space}%
\providecommand \EOS [0]{\spacefactor3000\relax}%
\providecommand \BibitemShut  [1]{\csname bibitem#1\endcsname}%
\let\auto@bib@innerbib\@empty
\bibitem [{\citenamefont {F\"orstner}\ \emph {et~al.}(2003)\citenamefont
  {F\"orstner}, \citenamefont {Weber}, \citenamefont {Danckwerts},\ and\
  \citenamefont {Knorr}}]{Foerstner2003}%
  \BibitemOpen
  \bibfield  {author} {\bibinfo {author} {\bibfnamefont {J.}~\bibnamefont
  {F\"orstner}}, \bibinfo {author} {\bibfnamefont {C.}~\bibnamefont {Weber}},
  \bibinfo {author} {\bibfnamefont {J.}~\bibnamefont {Danckwerts}}, \ and\
  \bibinfo {author} {\bibfnamefont {A.}~\bibnamefont {Knorr}},\ }\href
  {\doibase 10.1103/PhysRevLett.91.127401} {\bibfield  {journal} {\bibinfo
  {journal} {Phys. Rev. Lett.}\ }\textbf {\bibinfo {volume} {91}},\ \bibinfo
  {pages} {127401} (\bibinfo {year} {2003})}\BibitemShut {NoStop}%
\bibitem [{\citenamefont {Vagov}\ \emph {et~al.}(2004)\citenamefont {Vagov},
  \citenamefont {Axt}, \citenamefont {Kuhn}, \citenamefont {Langbein},
  \citenamefont {Borri},\ and\ \citenamefont {Woggon}}]{Vagov2004}%
  \BibitemOpen
  \bibfield  {author} {\bibinfo {author} {\bibfnamefont {A.}~\bibnamefont
  {Vagov}}, \bibinfo {author} {\bibfnamefont {V.~M.}\ \bibnamefont {Axt}},
  \bibinfo {author} {\bibfnamefont {T.}~\bibnamefont {Kuhn}}, \bibinfo {author}
  {\bibfnamefont {W.}~\bibnamefont {Langbein}}, \bibinfo {author}
  {\bibfnamefont {P.}~\bibnamefont {Borri}}, \ and\ \bibinfo {author}
  {\bibfnamefont {U.}~\bibnamefont {Woggon}},\ }\href {\doibase
  10.1103/PhysRevB.70.201305} {\bibfield  {journal} {\bibinfo  {journal} {Phys.
  Rev. B}\ }\textbf {\bibinfo {volume} {70}},\ \bibinfo {pages} {201305}
  (\bibinfo {year} {2004})}\BibitemShut {NoStop}%
\bibitem [{\citenamefont {Thorwart}\ \emph {et~al.}(2005)\citenamefont
  {Thorwart}, \citenamefont {Eckel},\ and\ \citenamefont
  {Mucciolo}}]{Thorwart2005}%
  \BibitemOpen
  \bibfield  {author} {\bibinfo {author} {\bibfnamefont {M.}~\bibnamefont
  {Thorwart}}, \bibinfo {author} {\bibfnamefont {J.}~\bibnamefont {Eckel}}, \
  and\ \bibinfo {author} {\bibfnamefont {E.~R.}\ \bibnamefont {Mucciolo}},\
  }\href {\doibase 10.1103/PhysRevB.72.235320} {\bibfield  {journal} {\bibinfo
  {journal} {Phys. Rev. B}\ }\textbf {\bibinfo {volume} {72}},\ \bibinfo
  {pages} {235320} (\bibinfo {year} {2005})}\BibitemShut {NoStop}%
\bibitem [{\citenamefont {Vagov}\ \emph {et~al.}(2007)\citenamefont {Vagov},
  \citenamefont {Croitoru}, \citenamefont {Axt}, \citenamefont {Kuhn},\ and\
  \citenamefont {Peeters}}]{Vagov2007}%
  \BibitemOpen
  \bibfield  {author} {\bibinfo {author} {\bibfnamefont {A.}~\bibnamefont
  {Vagov}}, \bibinfo {author} {\bibfnamefont {M.~D.}\ \bibnamefont {Croitoru}},
  \bibinfo {author} {\bibfnamefont {V.~M.}\ \bibnamefont {Axt}}, \bibinfo
  {author} {\bibfnamefont {T.}~\bibnamefont {Kuhn}}, \ and\ \bibinfo {author}
  {\bibfnamefont {F.~M.}\ \bibnamefont {Peeters}},\ }\href {\doibase
  10.1103/PhysRevLett.98.227403} {\bibfield  {journal} {\bibinfo  {journal}
  {Phys. Rev. Lett.}\ }\textbf {\bibinfo {volume} {98}},\ \bibinfo {pages}
  {227403} (\bibinfo {year} {2007})}\BibitemShut {NoStop}%
\bibitem [{\citenamefont {Ramsay}\ \emph
  {et~al.}(2010{\natexlab{a}})\citenamefont {Ramsay}, \citenamefont {Gopal},
  \citenamefont {Gauger}, \citenamefont {Nazir}, \citenamefont {Lovett},
  \citenamefont {Fox},\ and\ \citenamefont {Skolnick}}]{Ramsay2010a}%
  \BibitemOpen
  \bibfield  {author} {\bibinfo {author} {\bibfnamefont {A.~J.}\ \bibnamefont
  {Ramsay}}, \bibinfo {author} {\bibfnamefont {A.~V.}\ \bibnamefont {Gopal}},
  \bibinfo {author} {\bibfnamefont {E.~M.}\ \bibnamefont {Gauger}}, \bibinfo
  {author} {\bibfnamefont {A.}~\bibnamefont {Nazir}}, \bibinfo {author}
  {\bibfnamefont {B.~W.}\ \bibnamefont {Lovett}}, \bibinfo {author}
  {\bibfnamefont {A.~M.}\ \bibnamefont {Fox}}, \ and\ \bibinfo {author}
  {\bibfnamefont {M.~S.}\ \bibnamefont {Skolnick}},\ }\href {\doibase
  10.1103/PhysRevLett.104.017402} {\bibfield  {journal} {\bibinfo  {journal}
  {Phys. Rev. Lett.}\ }\textbf {\bibinfo {volume} {104}},\ \bibinfo {pages}
  {017402} (\bibinfo {year} {2010}{\natexlab{a}})}\BibitemShut {NoStop}%
\bibitem [{\citenamefont {Kaer}\ \emph {et~al.}(2010)\citenamefont {Kaer},
  \citenamefont {Nielsen}, \citenamefont {Lodahl}, \citenamefont {Jauho},\ and\
  \citenamefont {M\o{}rk}}]{Kaer2010}%
  \BibitemOpen
  \bibfield  {author} {\bibinfo {author} {\bibfnamefont {P.}~\bibnamefont
  {Kaer}}, \bibinfo {author} {\bibfnamefont {T.~R.}\ \bibnamefont {Nielsen}},
  \bibinfo {author} {\bibfnamefont {P.}~\bibnamefont {Lodahl}}, \bibinfo
  {author} {\bibfnamefont {A.-P.}\ \bibnamefont {Jauho}}, \ and\ \bibinfo
  {author} {\bibfnamefont {J.}~\bibnamefont {M\o{}rk}},\ }\href {\doibase
  10.1103/PhysRevLett.104.157401} {\bibfield  {journal} {\bibinfo  {journal}
  {Phys. Rev. Lett.}\ }\textbf {\bibinfo {volume} {104}},\ \bibinfo {pages}
  {157401} (\bibinfo {year} {2010})}\BibitemShut {NoStop}%
\bibitem [{\citenamefont {Michler}\ \emph {et~al.}(2000)\citenamefont
  {Michler}, \citenamefont {Kiraz}, \citenamefont {Becher}, \citenamefont
  {Schoenfeld}, \citenamefont {Petroff}, \citenamefont {Zhang}, \citenamefont
  {Hu},\ and\ \citenamefont {Imamoglu}}]{Michler2000}%
  \BibitemOpen
  \bibfield  {author} {\bibinfo {author} {\bibfnamefont {P.}~\bibnamefont
  {Michler}}, \bibinfo {author} {\bibfnamefont {A.}~\bibnamefont {Kiraz}},
  \bibinfo {author} {\bibfnamefont {C.}~\bibnamefont {Becher}}, \bibinfo
  {author} {\bibfnamefont {W.~V.}\ \bibnamefont {Schoenfeld}}, \bibinfo
  {author} {\bibfnamefont {P.~M.}\ \bibnamefont {Petroff}}, \bibinfo {author}
  {\bibfnamefont {L.}~\bibnamefont {Zhang}}, \bibinfo {author} {\bibfnamefont
  {E.}~\bibnamefont {Hu}}, \ and\ \bibinfo {author} {\bibfnamefont
  {A.}~\bibnamefont {Imamoglu}},\ }\href {\doibase
  10.1126/science.290.5500.2282} {\bibfield  {journal} {\bibinfo  {journal}
  {Science}\ }\textbf {\bibinfo {volume} {290}},\ \bibinfo {pages} {2282}
  (\bibinfo {year} {2000})}\BibitemShut {NoStop}%
\bibitem [{\citenamefont {He}\ \emph {et~al.}(2013)\citenamefont {He},
  \citenamefont {He}, \citenamefont {Wei}, \citenamefont {Wu}, \citenamefont
  {Atat{\"u}re}, \citenamefont {Schneider}, \citenamefont {H{\"o}fling},
  \citenamefont {Kamp}, \citenamefont {Lu},\ and\ \citenamefont
  {Pan}}]{He2013}%
  \BibitemOpen
  \bibfield  {author} {\bibinfo {author} {\bibfnamefont {Y.-M.}\ \bibnamefont
  {He}}, \bibinfo {author} {\bibfnamefont {Y.}~\bibnamefont {He}}, \bibinfo
  {author} {\bibfnamefont {Y.-J.}\ \bibnamefont {Wei}}, \bibinfo {author}
  {\bibfnamefont {D.}~\bibnamefont {Wu}}, \bibinfo {author} {\bibfnamefont
  {M.}~\bibnamefont {Atat{\"u}re}}, \bibinfo {author} {\bibfnamefont
  {C.}~\bibnamefont {Schneider}}, \bibinfo {author} {\bibfnamefont
  {S.}~\bibnamefont {H{\"o}fling}}, \bibinfo {author} {\bibfnamefont
  {M.}~\bibnamefont {Kamp}}, \bibinfo {author} {\bibfnamefont {C.-Y.}\
  \bibnamefont {Lu}}, \ and\ \bibinfo {author} {\bibfnamefont {J.-W.}\
  \bibnamefont {Pan}},\ }\href {http://dx.doi.org/10.1038/nnano.2012.262}
  {\bibfield  {journal} {\bibinfo  {journal} {Nature Nanotechnology}\ }\textbf
  {\bibinfo {volume} {8}},\ \bibinfo {pages} {213 EP } (\bibinfo {year}
  {2013})}\BibitemShut {NoStop}%
\bibitem [{\citenamefont {Somaschi}\ \emph {et~al.}(2016)\citenamefont
  {Somaschi}, \citenamefont {Giesz}, \citenamefont {De~Santis}, \citenamefont
  {Loredo}, \citenamefont {Almeida}, \citenamefont {Hornecker}, \citenamefont
  {Portalupi}, \citenamefont {Grange}, \citenamefont {Ant{\'o}n}, \citenamefont
  {Demory}, \citenamefont {G{\'o}mez}, \citenamefont {Sagnes}, \citenamefont
  {Lanzillotti-Kimura}, \citenamefont {Lema{\'i}tre}, \citenamefont {Auffeves},
  \citenamefont {White}, \citenamefont {Lanco},\ and\ \citenamefont
  {Senellart}}]{Somaschi2016}%
  \BibitemOpen
  \bibfield  {author} {\bibinfo {author} {\bibfnamefont {N.}~\bibnamefont
  {Somaschi}}, \bibinfo {author} {\bibfnamefont {V.}~\bibnamefont {Giesz}},
  \bibinfo {author} {\bibfnamefont {L.}~\bibnamefont {De~Santis}}, \bibinfo
  {author} {\bibfnamefont {J.~C.}\ \bibnamefont {Loredo}}, \bibinfo {author}
  {\bibfnamefont {M.~P.}\ \bibnamefont {Almeida}}, \bibinfo {author}
  {\bibfnamefont {G.}~\bibnamefont {Hornecker}}, \bibinfo {author}
  {\bibfnamefont {S.~L.}\ \bibnamefont {Portalupi}}, \bibinfo {author}
  {\bibfnamefont {T.}~\bibnamefont {Grange}}, \bibinfo {author} {\bibfnamefont
  {C.}~\bibnamefont {Ant{\'o}n}}, \bibinfo {author} {\bibfnamefont
  {J.}~\bibnamefont {Demory}}, \bibinfo {author} {\bibfnamefont
  {C.}~\bibnamefont {G{\'o}mez}}, \bibinfo {author} {\bibfnamefont
  {I.}~\bibnamefont {Sagnes}}, \bibinfo {author} {\bibfnamefont {N.~D.}\
  \bibnamefont {Lanzillotti-Kimura}}, \bibinfo {author} {\bibfnamefont
  {A.}~\bibnamefont {Lema{\'i}tre}}, \bibinfo {author} {\bibfnamefont
  {A.}~\bibnamefont {Auffeves}}, \bibinfo {author} {\bibfnamefont {A.~G.}\
  \bibnamefont {White}}, \bibinfo {author} {\bibfnamefont {L.}~\bibnamefont
  {Lanco}}, \ and\ \bibinfo {author} {\bibfnamefont {P.}~\bibnamefont
  {Senellart}},\ }\href {http://dx.doi.org/10.1038/nphoton.2016.23} {\bibfield
  {journal} {\bibinfo  {journal} {Nature Photonics}\ }\textbf {\bibinfo
  {volume} {10}},\ \bibinfo {pages} {340} (\bibinfo {year} {2016})}\BibitemShut
  {NoStop}%
\bibitem [{\citenamefont {Stevenson}\ \emph {et~al.}(2006)\citenamefont
  {Stevenson}, \citenamefont {Young}, \citenamefont {Atkinson}, \citenamefont
  {Cooper}, \citenamefont {Ritchie},\ and\ \citenamefont
  {Shields}}]{Stevenson2006}%
  \BibitemOpen
  \bibfield  {author} {\bibinfo {author} {\bibfnamefont {R.~M.}\ \bibnamefont
  {Stevenson}}, \bibinfo {author} {\bibfnamefont {R.~J.}\ \bibnamefont
  {Young}}, \bibinfo {author} {\bibfnamefont {P.}~\bibnamefont {Atkinson}},
  \bibinfo {author} {\bibfnamefont {K.}~\bibnamefont {Cooper}}, \bibinfo
  {author} {\bibfnamefont {D.~A.}\ \bibnamefont {Ritchie}}, \ and\ \bibinfo
  {author} {\bibfnamefont {A.~J.}\ \bibnamefont {Shields}},\ }\href
  {http://dx.doi.org/10.1038/nature04446} {\bibfield  {journal} {\bibinfo
  {journal} {Nature}\ }\textbf {\bibinfo {volume} {439}},\ \bibinfo {pages}
  {179 EP } (\bibinfo {year} {2006})}\BibitemShut {NoStop}%
\bibitem [{\citenamefont {Akopian}\ \emph {et~al.}(2006)\citenamefont
  {Akopian}, \citenamefont {Lindner}, \citenamefont {Poem}, \citenamefont
  {Berlatzky}, \citenamefont {Avron}, \citenamefont {Gershoni}, \citenamefont
  {Gerardot},\ and\ \citenamefont {Petroff}}]{Akopian2006}%
  \BibitemOpen
  \bibfield  {author} {\bibinfo {author} {\bibfnamefont {N.}~\bibnamefont
  {Akopian}}, \bibinfo {author} {\bibfnamefont {N.~H.}\ \bibnamefont
  {Lindner}}, \bibinfo {author} {\bibfnamefont {E.}~\bibnamefont {Poem}},
  \bibinfo {author} {\bibfnamefont {Y.}~\bibnamefont {Berlatzky}}, \bibinfo
  {author} {\bibfnamefont {J.}~\bibnamefont {Avron}}, \bibinfo {author}
  {\bibfnamefont {D.}~\bibnamefont {Gershoni}}, \bibinfo {author}
  {\bibfnamefont {B.~D.}\ \bibnamefont {Gerardot}}, \ and\ \bibinfo {author}
  {\bibfnamefont {P.~M.}\ \bibnamefont {Petroff}},\ }\href {\doibase
  10.1103/PhysRevLett.96.130501} {\bibfield  {journal} {\bibinfo  {journal}
  {Phys. Rev. Lett.}\ }\textbf {\bibinfo {volume} {96}},\ \bibinfo {pages}
  {130501} (\bibinfo {year} {2006})}\BibitemShut {NoStop}%
\bibitem [{\citenamefont {Hafenbrak}\ \emph {et~al.}(2007)\citenamefont
  {Hafenbrak}, \citenamefont {Ulrich}, \citenamefont {Michler}, \citenamefont
  {Wang}, \citenamefont {Rastelli},\ and\ \citenamefont
  {Schmidt}}]{Hafenbrak2007}%
  \BibitemOpen
  \bibfield  {author} {\bibinfo {author} {\bibfnamefont {R.}~\bibnamefont
  {Hafenbrak}}, \bibinfo {author} {\bibfnamefont {S.~M.}\ \bibnamefont
  {Ulrich}}, \bibinfo {author} {\bibfnamefont {P.}~\bibnamefont {Michler}},
  \bibinfo {author} {\bibfnamefont {L.}~\bibnamefont {Wang}}, \bibinfo {author}
  {\bibfnamefont {A.}~\bibnamefont {Rastelli}}, \ and\ \bibinfo {author}
  {\bibfnamefont {O.~G.}\ \bibnamefont {Schmidt}},\ }\href
  {http://stacks.iop.org/1367-2630/9/i=9/a=315} {\bibfield  {journal} {\bibinfo
   {journal} {New Journal of Physics}\ }\textbf {\bibinfo {volume} {9}},\
  \bibinfo {pages} {315} (\bibinfo {year} {2007})}\BibitemShut {NoStop}%
\bibitem [{\citenamefont {Dousse}\ \emph {et~al.}(2010)\citenamefont {Dousse},
  \citenamefont {Suffczynski}, \citenamefont {Beveratos}, \citenamefont
  {Krebs}, \citenamefont {Lema{\^i}tre}, \citenamefont {Sagnes}, \citenamefont
  {Bloch}, \citenamefont {Voisin},\ and\ \citenamefont
  {Senellart}}]{Dousse2010}%
  \BibitemOpen
  \bibfield  {author} {\bibinfo {author} {\bibfnamefont {A.}~\bibnamefont
  {Dousse}}, \bibinfo {author} {\bibfnamefont {J.}~\bibnamefont {Suffczynski}},
  \bibinfo {author} {\bibfnamefont {A.}~\bibnamefont {Beveratos}}, \bibinfo
  {author} {\bibfnamefont {O.}~\bibnamefont {Krebs}}, \bibinfo {author}
  {\bibfnamefont {A.}~\bibnamefont {Lema{\^i}tre}}, \bibinfo {author}
  {\bibfnamefont {I.}~\bibnamefont {Sagnes}}, \bibinfo {author} {\bibfnamefont
  {J.}~\bibnamefont {Bloch}}, \bibinfo {author} {\bibfnamefont
  {P.}~\bibnamefont {Voisin}}, \ and\ \bibinfo {author} {\bibfnamefont
  {P.}~\bibnamefont {Senellart}},\ }\href
  {http://dx.doi.org/10.1038/nature09148} {\bibfield  {journal} {\bibinfo
  {journal} {Nature}\ }\textbf {\bibinfo {volume} {466}},\ \bibinfo {pages}
  {217 EP } (\bibinfo {year} {2010})}\BibitemShut {NoStop}%
\bibitem [{\citenamefont {M{\"u}ller}\ \emph {et~al.}(2014)\citenamefont
  {M{\"u}ller}, \citenamefont {Bounouar}, \citenamefont {J{\"o}ns},
  \citenamefont {Gl{\"a}ssl},\ and\ \citenamefont {Michler}}]{Mueller2014}%
  \BibitemOpen
  \bibfield  {author} {\bibinfo {author} {\bibfnamefont {M.}~\bibnamefont
  {M{\"u}ller}}, \bibinfo {author} {\bibfnamefont {S.}~\bibnamefont
  {Bounouar}}, \bibinfo {author} {\bibfnamefont {K.~D.}\ \bibnamefont
  {J{\"o}ns}}, \bibinfo {author} {\bibfnamefont {M.}~\bibnamefont
  {Gl{\"a}ssl}}, \ and\ \bibinfo {author} {\bibfnamefont {P.}~\bibnamefont
  {Michler}},\ }\href {http://dx.doi.org/10.1038/nphoton.2013.377} {\bibfield
  {journal} {\bibinfo  {journal} {Nature Photonics}\ }\textbf {\bibinfo
  {volume} {8}},\ \bibinfo {pages} {224 EP } (\bibinfo {year}
  {2014})}\BibitemShut {NoStop}%
\bibitem [{\citenamefont {Besombes}\ \emph {et~al.}(2001)\citenamefont
  {Besombes}, \citenamefont {Kheng}, \citenamefont {Marsal},\ and\
  \citenamefont {Mariette}}]{Besombes2001}%
  \BibitemOpen
  \bibfield  {author} {\bibinfo {author} {\bibfnamefont {L.}~\bibnamefont
  {Besombes}}, \bibinfo {author} {\bibfnamefont {K.}~\bibnamefont {Kheng}},
  \bibinfo {author} {\bibfnamefont {L.}~\bibnamefont {Marsal}}, \ and\ \bibinfo
  {author} {\bibfnamefont {H.}~\bibnamefont {Mariette}},\ }\href {\doibase
  10.1103/PhysRevB.63.155307} {\bibfield  {journal} {\bibinfo  {journal} {Phys.
  Rev. B}\ }\textbf {\bibinfo {volume} {63}},\ \bibinfo {pages} {155307}
  (\bibinfo {year} {2001})}\BibitemShut {NoStop}%
\bibitem [{\citenamefont {Borri}\ \emph {et~al.}(2001)\citenamefont {Borri},
  \citenamefont {Langbein}, \citenamefont {Schneider}, \citenamefont {Woggon},
  \citenamefont {Sellin}, \citenamefont {Ouyang},\ and\ \citenamefont
  {Bimberg}}]{Borri2001}%
  \BibitemOpen
  \bibfield  {author} {\bibinfo {author} {\bibfnamefont {P.}~\bibnamefont
  {Borri}}, \bibinfo {author} {\bibfnamefont {W.}~\bibnamefont {Langbein}},
  \bibinfo {author} {\bibfnamefont {S.}~\bibnamefont {Schneider}}, \bibinfo
  {author} {\bibfnamefont {U.}~\bibnamefont {Woggon}}, \bibinfo {author}
  {\bibfnamefont {R.~L.}\ \bibnamefont {Sellin}}, \bibinfo {author}
  {\bibfnamefont {D.}~\bibnamefont {Ouyang}}, \ and\ \bibinfo {author}
  {\bibfnamefont {D.}~\bibnamefont {Bimberg}},\ }\href {\doibase
  10.1103/PhysRevLett.87.157401} {\bibfield  {journal} {\bibinfo  {journal}
  {Phys. Rev. Lett.}\ }\textbf {\bibinfo {volume} {87}},\ \bibinfo {pages}
  {157401} (\bibinfo {year} {2001})}\BibitemShut {NoStop}%
\bibitem [{\citenamefont {Krummheuer}\ \emph {et~al.}(2002)\citenamefont
  {Krummheuer}, \citenamefont {Axt},\ and\ \citenamefont
  {Kuhn}}]{Krummheuer2002}%
  \BibitemOpen
  \bibfield  {author} {\bibinfo {author} {\bibfnamefont {B.}~\bibnamefont
  {Krummheuer}}, \bibinfo {author} {\bibfnamefont {V.~M.}\ \bibnamefont {Axt}},
  \ and\ \bibinfo {author} {\bibfnamefont {T.}~\bibnamefont {Kuhn}},\ }\href
  {\doibase 10.1103/PhysRevB.65.195313} {\bibfield  {journal} {\bibinfo
  {journal} {Phys. Rev. B}\ }\textbf {\bibinfo {volume} {65}},\ \bibinfo
  {pages} {195313} (\bibinfo {year} {2002})}\BibitemShut {NoStop}%
\bibitem [{\citenamefont {Ulrich}\ \emph {et~al.}(2011)\citenamefont {Ulrich},
  \citenamefont {Ates}, \citenamefont {Reitzenstein}, \citenamefont
  {L\"offler}, \citenamefont {Forchel},\ and\ \citenamefont
  {Michler}}]{Ulrich2011}%
  \BibitemOpen
  \bibfield  {author} {\bibinfo {author} {\bibfnamefont {S.~M.}\ \bibnamefont
  {Ulrich}}, \bibinfo {author} {\bibfnamefont {S.}~\bibnamefont {Ates}},
  \bibinfo {author} {\bibfnamefont {S.}~\bibnamefont {Reitzenstein}}, \bibinfo
  {author} {\bibfnamefont {A.}~\bibnamefont {L\"offler}}, \bibinfo {author}
  {\bibfnamefont {A.}~\bibnamefont {Forchel}}, \ and\ \bibinfo {author}
  {\bibfnamefont {P.}~\bibnamefont {Michler}},\ }\href {\doibase
  10.1103/PhysRevLett.106.247402} {\bibfield  {journal} {\bibinfo  {journal}
  {Phys. Rev. Lett.}\ }\textbf {\bibinfo {volume} {106}},\ \bibinfo {pages}
  {247402} (\bibinfo {year} {2011})}\BibitemShut {NoStop}%
\bibitem [{\citenamefont {Roy}\ and\ \citenamefont
  {Hughes}(2011{\natexlab{a}})}]{Roy2011a}%
  \BibitemOpen
  \bibfield  {author} {\bibinfo {author} {\bibfnamefont {C.}~\bibnamefont
  {Roy}}\ and\ \bibinfo {author} {\bibfnamefont {S.}~\bibnamefont {Hughes}},\
  }\href {\doibase 10.1103/PhysRevLett.106.247403} {\bibfield  {journal}
  {\bibinfo  {journal} {Phys. Rev. Lett.}\ }\textbf {\bibinfo {volume} {106}},\
  \bibinfo {pages} {247403} (\bibinfo {year} {2011}{\natexlab{a}})}\BibitemShut
  {NoStop}%
\bibitem [{\citenamefont {Roy}\ and\ \citenamefont
  {Hughes}(2011{\natexlab{b}})}]{Roy2011b}%
  \BibitemOpen
  \bibfield  {author} {\bibinfo {author} {\bibfnamefont {C.}~\bibnamefont
  {Roy}}\ and\ \bibinfo {author} {\bibfnamefont {S.}~\bibnamefont {Hughes}},\
  }\href {\doibase 10.1103/PhysRevX.1.021009} {\bibfield  {journal} {\bibinfo
  {journal} {Phys. Rev. X}\ }\textbf {\bibinfo {volume} {1}},\ \bibinfo {pages}
  {021009} (\bibinfo {year} {2011}{\natexlab{b}})}\BibitemShut {NoStop}%
\bibitem [{\citenamefont {Harsij}\ \emph {et~al.}(2012)\citenamefont {Harsij},
  \citenamefont {Bagheri~Harouni}, \citenamefont {Roknizadeh},\ and\
  \citenamefont {Naderi}}]{Harsij2012}%
  \BibitemOpen
  \bibfield  {author} {\bibinfo {author} {\bibfnamefont {Z.}~\bibnamefont
  {Harsij}}, \bibinfo {author} {\bibfnamefont {M.}~\bibnamefont
  {Bagheri~Harouni}}, \bibinfo {author} {\bibfnamefont {R.}~\bibnamefont
  {Roknizadeh}}, \ and\ \bibinfo {author} {\bibfnamefont {M.~H.}\ \bibnamefont
  {Naderi}},\ }\href {\doibase 10.1103/PhysRevA.86.063803} {\bibfield
  {journal} {\bibinfo  {journal} {Phys. Rev. A}\ }\textbf {\bibinfo {volume}
  {86}},\ \bibinfo {pages} {063803} (\bibinfo {year} {2012})}\BibitemShut
  {NoStop}%
\bibitem [{\citenamefont {Roy}\ and\ \citenamefont {Hughes}(2012)}]{Roy2012}%
  \BibitemOpen
  \bibfield  {author} {\bibinfo {author} {\bibfnamefont {C.}~\bibnamefont
  {Roy}}\ and\ \bibinfo {author} {\bibfnamefont {S.}~\bibnamefont {Hughes}},\
  }\href {\doibase 10.1103/PhysRevB.85.115309} {\bibfield  {journal} {\bibinfo
  {journal} {Phys. Rev. B}\ }\textbf {\bibinfo {volume} {85}},\ \bibinfo
  {pages} {115309} (\bibinfo {year} {2012})}\BibitemShut {NoStop}%
\bibitem [{\citenamefont {McCutcheon}\ and\ \citenamefont
  {Nazir}(2013)}]{McCutcheon2013}%
  \BibitemOpen
  \bibfield  {author} {\bibinfo {author} {\bibfnamefont {D.~P.~S.}\
  \bibnamefont {McCutcheon}}\ and\ \bibinfo {author} {\bibfnamefont
  {A.}~\bibnamefont {Nazir}},\ }\href {\doibase 10.1103/PhysRevLett.110.217401}
  {\bibfield  {journal} {\bibinfo  {journal} {Phys. Rev. Lett.}\ }\textbf
  {\bibinfo {volume} {110}},\ \bibinfo {pages} {217401} (\bibinfo {year}
  {2013})}\BibitemShut {NoStop}%
\bibitem [{\citenamefont {Roy-Choudhury}\ and\ \citenamefont
  {Hughes}(2015)}]{Roy-Choudhury2015}%
  \BibitemOpen
  \bibfield  {author} {\bibinfo {author} {\bibfnamefont {K.}~\bibnamefont
  {Roy-Choudhury}}\ and\ \bibinfo {author} {\bibfnamefont {S.}~\bibnamefont
  {Hughes}},\ }\href {\doibase 10.1103/PhysRevB.92.205406} {\bibfield
  {journal} {\bibinfo  {journal} {Phys. Rev. B}\ }\textbf {\bibinfo {volume}
  {92}},\ \bibinfo {pages} {205406} (\bibinfo {year} {2015})}\BibitemShut
  {NoStop}%
\bibitem [{\citenamefont {McCutcheon}(2016)}]{McCutcheon2016}%
  \BibitemOpen
  \bibfield  {author} {\bibinfo {author} {\bibfnamefont {D.~P.~S.}\
  \bibnamefont {McCutcheon}},\ }\href {\doibase 10.1103/PhysRevA.93.022119}
  {\bibfield  {journal} {\bibinfo  {journal} {Phys. Rev. A}\ }\textbf {\bibinfo
  {volume} {93}},\ \bibinfo {pages} {022119} (\bibinfo {year}
  {2016})}\BibitemShut {NoStop}%
\bibitem [{\citenamefont {Hanbury~Brown}\ and\ \citenamefont
  {Twiss}(1956)}]{Hanbury1956}%
  \BibitemOpen
  \bibfield  {author} {\bibinfo {author} {\bibfnamefont {R.}~\bibnamefont
  {Hanbury~Brown}}\ and\ \bibinfo {author} {\bibfnamefont {R.~Q.}\ \bibnamefont
  {Twiss}},\ }\href {http://dx.doi.org/10.1038/1781046a0} {\bibfield  {journal}
  {\bibinfo  {journal} {Nature}\ }\textbf {\bibinfo {volume} {178}},\ \bibinfo
  {pages} {1046} (\bibinfo {year} {1956})}\BibitemShut {NoStop}%
\bibitem [{\citenamefont {Glauber}(2007)}]{Glauber2007}%
  \BibitemOpen
  \bibfield  {author} {\bibinfo {author} {\bibfnamefont {R.~J.}\ \bibnamefont
  {Glauber}},\ }\href@noop {} {\emph {\bibinfo {title} {Quantum Theory of
  Optical Coherence: Selected Papers and Lectures}}}\ (\bibinfo  {publisher}
  {Wiley},\ \bibinfo {year} {2007})\BibitemShut {NoStop}%
\bibitem [{\citenamefont {Santori}\ \emph {et~al.}(2002)\citenamefont
  {Santori}, \citenamefont {Fattal}, \citenamefont {Vuckovic}, \citenamefont
  {Solomon},\ and\ \citenamefont {Yamamoto}}]{Santori2002}%
  \BibitemOpen
  \bibfield  {author} {\bibinfo {author} {\bibfnamefont {C.}~\bibnamefont
  {Santori}}, \bibinfo {author} {\bibfnamefont {D.}~\bibnamefont {Fattal}},
  \bibinfo {author} {\bibfnamefont {J.}~\bibnamefont {Vuckovic}}, \bibinfo
  {author} {\bibfnamefont {G.~S.}\ \bibnamefont {Solomon}}, \ and\ \bibinfo
  {author} {\bibfnamefont {Y.}~\bibnamefont {Yamamoto}},\ }\href
  {http://dx.doi.org/10.1038/nature01086} {\bibfield  {journal} {\bibinfo
  {journal} {Nature}\ }\textbf {\bibinfo {volume} {419}},\ \bibinfo {pages}
  {594} (\bibinfo {year} {2002})}\BibitemShut {NoStop}%
\bibitem [{\citenamefont {Bentham}\ \emph {et~al.}(2016)\citenamefont
  {Bentham}, \citenamefont {Hallett}, \citenamefont {Prtljaga}, \citenamefont
  {Royall}, \citenamefont {Vaitiekus}, \citenamefont {Coles}, \citenamefont
  {Clarke}, \citenamefont {Fox}, \citenamefont {Skolnick}, \citenamefont
  {Itskevich},\ and\ \citenamefont {Wilson}}]{Bentham2016}%
  \BibitemOpen
  \bibfield  {author} {\bibinfo {author} {\bibfnamefont {C.}~\bibnamefont
  {Bentham}}, \bibinfo {author} {\bibfnamefont {D.}~\bibnamefont {Hallett}},
  \bibinfo {author} {\bibfnamefont {N.}~\bibnamefont {Prtljaga}}, \bibinfo
  {author} {\bibfnamefont {B.}~\bibnamefont {Royall}}, \bibinfo {author}
  {\bibfnamefont {D.}~\bibnamefont {Vaitiekus}}, \bibinfo {author}
  {\bibfnamefont {R.~J.}\ \bibnamefont {Coles}}, \bibinfo {author}
  {\bibfnamefont {E.}~\bibnamefont {Clarke}}, \bibinfo {author} {\bibfnamefont
  {A.~M.}\ \bibnamefont {Fox}}, \bibinfo {author} {\bibfnamefont {M.~S.}\
  \bibnamefont {Skolnick}}, \bibinfo {author} {\bibfnamefont {I.~E.}\
  \bibnamefont {Itskevich}}, \ and\ \bibinfo {author} {\bibfnamefont {L.~R.}\
  \bibnamefont {Wilson}},\ }\href {\doibase 10.1063/1.4965295} {\bibfield
  {journal} {\bibinfo  {journal} {Applied Physics Letters}\ }\textbf {\bibinfo
  {volume} {109}},\ \bibinfo {pages} {161101} (\bibinfo {year}
  {2016})}\BibitemShut {NoStop}%
\bibitem [{\citenamefont {Prtljaga}\ \emph {et~al.}(2016)\citenamefont
  {Prtljaga}, \citenamefont {Bentham}, \citenamefont {O'Hara}, \citenamefont
  {Royall}, \citenamefont {Clarke}, \citenamefont {Wilson}, \citenamefont
  {Skolnick},\ and\ \citenamefont {Fox}}]{Prtljaga2016}%
  \BibitemOpen
  \bibfield  {author} {\bibinfo {author} {\bibfnamefont {N.}~\bibnamefont
  {Prtljaga}}, \bibinfo {author} {\bibfnamefont {C.}~\bibnamefont {Bentham}},
  \bibinfo {author} {\bibfnamefont {J.}~\bibnamefont {O'Hara}}, \bibinfo
  {author} {\bibfnamefont {B.}~\bibnamefont {Royall}}, \bibinfo {author}
  {\bibfnamefont {E.}~\bibnamefont {Clarke}}, \bibinfo {author} {\bibfnamefont
  {L.~R.}\ \bibnamefont {Wilson}}, \bibinfo {author} {\bibfnamefont {M.~S.}\
  \bibnamefont {Skolnick}}, \ and\ \bibinfo {author} {\bibfnamefont {A.~M.}\
  \bibnamefont {Fox}},\ }\href {\doibase 10.1063/1.4954220} {\bibfield
  {journal} {\bibinfo  {journal} {Applied Physics Letters}\ }\textbf {\bibinfo
  {volume} {108}},\ \bibinfo {pages} {251101} (\bibinfo {year}
  {2016})}\BibitemShut {NoStop}%
\bibitem [{\citenamefont {Wei{\ss}}\ \emph {et~al.}(2016)\citenamefont
  {Wei{\ss}}, \citenamefont {Kapfinger}, \citenamefont {Reichert},
  \citenamefont {Finley}, \citenamefont {Wixforth}, \citenamefont {Kaniber},\
  and\ \citenamefont {Krenner}}]{Weiss2016}%
  \BibitemOpen
  \bibfield  {author} {\bibinfo {author} {\bibfnamefont {M.}~\bibnamefont
  {Wei{\ss}}}, \bibinfo {author} {\bibfnamefont {S.}~\bibnamefont {Kapfinger}},
  \bibinfo {author} {\bibfnamefont {T.}~\bibnamefont {Reichert}}, \bibinfo
  {author} {\bibfnamefont {J.~J.}\ \bibnamefont {Finley}}, \bibinfo {author}
  {\bibfnamefont {A.}~\bibnamefont {Wixforth}}, \bibinfo {author}
  {\bibfnamefont {M.}~\bibnamefont {Kaniber}}, \ and\ \bibinfo {author}
  {\bibfnamefont {H.~J.}\ \bibnamefont {Krenner}},\ }\href {\doibase
  10.1063/1.4959079} {\bibfield  {journal} {\bibinfo  {journal} {Applied
  Physics Letters}\ }\textbf {\bibinfo {volume} {109}},\ \bibinfo {pages}
  {033105} (\bibinfo {year} {2016})}\BibitemShut {NoStop}%
\bibitem [{\citenamefont {Ding}\ \emph {et~al.}(2016)\citenamefont {Ding},
  \citenamefont {He}, \citenamefont {Duan}, \citenamefont {Gregersen},
  \citenamefont {Chen}, \citenamefont {Unsleber}, \citenamefont {Maier},
  \citenamefont {Schneider}, \citenamefont {Kamp}, \citenamefont {H{\"o}fling},
  \citenamefont {Lu},\ and\ \citenamefont {Pan}}]{Ding2016}%
  \BibitemOpen
  \bibfield  {author} {\bibinfo {author} {\bibfnamefont {X.}~\bibnamefont
  {Ding}}, \bibinfo {author} {\bibfnamefont {Y.}~\bibnamefont {He}}, \bibinfo
  {author} {\bibfnamefont {Z.-C.}\ \bibnamefont {Duan}}, \bibinfo {author}
  {\bibfnamefont {N.}~\bibnamefont {Gregersen}}, \bibinfo {author}
  {\bibfnamefont {M.-C.}\ \bibnamefont {Chen}}, \bibinfo {author}
  {\bibfnamefont {S.}~\bibnamefont {Unsleber}}, \bibinfo {author}
  {\bibfnamefont {S.}~\bibnamefont {Maier}}, \bibinfo {author} {\bibfnamefont
  {C.}~\bibnamefont {Schneider}}, \bibinfo {author} {\bibfnamefont
  {M.}~\bibnamefont {Kamp}}, \bibinfo {author} {\bibfnamefont {S.}~\bibnamefont
  {H{\"o}fling}}, \bibinfo {author} {\bibfnamefont {C.-Y.}\ \bibnamefont {Lu}},
  \ and\ \bibinfo {author} {\bibfnamefont {J.-W.}\ \bibnamefont {Pan}},\ }\href
  {\doibase 10.1103/PhysRevLett.116.020401} {\bibfield  {journal} {\bibinfo
  {journal} {Phys. Rev. Lett.}\ }\textbf {\bibinfo {volume} {116}},\ \bibinfo
  {pages} {020401} (\bibinfo {year} {2016})}\BibitemShut {NoStop}%
\bibitem [{\citenamefont {Schweickert}\ \emph {et~al.}(2018)\citenamefont
  {Schweickert}, \citenamefont {Jöns}, \citenamefont {Zeuner}, \citenamefont
  {da~Silva}, \citenamefont {Huang}, \citenamefont {Lettner}, \citenamefont
  {Reindl}, \citenamefont {Zichi}, \citenamefont {Trotta}, \citenamefont
  {Rastelli},\ and\ \citenamefont {Zwiller}}]{Schweickert2017}%
  \BibitemOpen
  \bibfield  {author} {\bibinfo {author} {\bibfnamefont {L.}~\bibnamefont
  {Schweickert}}, \bibinfo {author} {\bibfnamefont {K.~D.}\ \bibnamefont
  {Jöns}}, \bibinfo {author} {\bibfnamefont {K.~D.}\ \bibnamefont {Zeuner}},
  \bibinfo {author} {\bibfnamefont {S.~F.~C.}\ \bibnamefont {da~Silva}},
  \bibinfo {author} {\bibfnamefont {H.}~\bibnamefont {Huang}}, \bibinfo
  {author} {\bibfnamefont {T.}~\bibnamefont {Lettner}}, \bibinfo {author}
  {\bibfnamefont {M.}~\bibnamefont {Reindl}}, \bibinfo {author} {\bibfnamefont
  {J.}~\bibnamefont {Zichi}}, \bibinfo {author} {\bibfnamefont
  {R.}~\bibnamefont {Trotta}}, \bibinfo {author} {\bibfnamefont
  {A.}~\bibnamefont {Rastelli}}, \ and\ \bibinfo {author} {\bibfnamefont
  {V.}~\bibnamefont {Zwiller}},\ }\href {\doibase 10.1063/1.5020038} {\bibfield
   {journal} {\bibinfo  {journal} {Applied Physics Letters}\ }\textbf {\bibinfo
  {volume} {112}},\ \bibinfo {pages} {093106} (\bibinfo {year} {2018})},\
  \Eprint {http://arxiv.org/abs/https://doi.org/10.1063/1.5020038}
  {https://doi.org/10.1063/1.5020038} \BibitemShut {NoStop}%
\bibitem [{\citenamefont {Mu{\~n}oz}\ \emph {et~al.}(2014)\citenamefont
  {Mu{\~n}oz}, \citenamefont {del Valle}, \citenamefont {Tudela}, \citenamefont
  {M{\"u}ller}, \citenamefont {Lichtmannecker}, \citenamefont {Kaniber},
  \citenamefont {Tejedor}, \citenamefont {Finley},\ and\ \citenamefont
  {Laussy}}]{Munoz2014}%
  \BibitemOpen
  \bibfield  {author} {\bibinfo {author} {\bibfnamefont {C.~S.}\ \bibnamefont
  {Mu{\~n}oz}}, \bibinfo {author} {\bibfnamefont {E.}~\bibnamefont {del
  Valle}}, \bibinfo {author} {\bibfnamefont {A.~G.}\ \bibnamefont {Tudela}},
  \bibinfo {author} {\bibfnamefont {K.}~\bibnamefont {M{\"u}ller}}, \bibinfo
  {author} {\bibfnamefont {S.}~\bibnamefont {Lichtmannecker}}, \bibinfo
  {author} {\bibfnamefont {M.}~\bibnamefont {Kaniber}}, \bibinfo {author}
  {\bibfnamefont {C.}~\bibnamefont {Tejedor}}, \bibinfo {author} {\bibfnamefont
  {J.~J.}\ \bibnamefont {Finley}}, \ and\ \bibinfo {author} {\bibfnamefont
  {F.~P.}\ \bibnamefont {Laussy}},\ }\href
  {http://dx.doi.org/10.1038/nphoton.2014.114} {\bibfield  {journal} {\bibinfo
  {journal} {Nature Photonics}\ }\textbf {\bibinfo {volume} {8}},\ \bibinfo
  {pages} {550 EP } (\bibinfo {year} {2014})}\BibitemShut {NoStop}%
\bibitem [{\citenamefont {Carmichael}(1993)}]{Carmichael1993}%
  \BibitemOpen
  \bibfield  {author} {\bibinfo {author} {\bibfnamefont {H.}~\bibnamefont
  {Carmichael}},\ }\href@noop {} {\emph {\bibinfo {title} {An open systems
  approach to quantum optics}}}\ (\bibinfo  {publisher} {Springer},\ \bibinfo
  {year} {1993})\BibitemShut {NoStop}%
\bibitem [{\citenamefont {Makri}\ and\ \citenamefont
  {Makarov}(1995{\natexlab{a}})}]{Makri1995a}%
  \BibitemOpen
  \bibfield  {author} {\bibinfo {author} {\bibfnamefont {N.}~\bibnamefont
  {Makri}}\ and\ \bibinfo {author} {\bibfnamefont {D.~E.}\ \bibnamefont
  {Makarov}},\ }\href {\doibase 10.1063/1.469508} {\bibfield  {journal}
  {\bibinfo  {journal} {The Journal of Chemical Physics}\ }\textbf {\bibinfo
  {volume} {102}},\ \bibinfo {pages} {4600} (\bibinfo {year}
  {1995}{\natexlab{a}})},\ \Eprint
  {http://arxiv.org/abs/https://doi.org/10.1063/1.469508}
  {https://doi.org/10.1063/1.469508} \BibitemShut {NoStop}%
\bibitem [{\citenamefont {Makri}\ and\ \citenamefont
  {Makarov}(1995{\natexlab{b}})}]{Makri1995b}%
  \BibitemOpen
  \bibfield  {author} {\bibinfo {author} {\bibfnamefont {N.}~\bibnamefont
  {Makri}}\ and\ \bibinfo {author} {\bibfnamefont {D.~E.}\ \bibnamefont
  {Makarov}},\ }\href {\doibase 10.1063/1.469509} {\bibfield  {journal}
  {\bibinfo  {journal} {The Journal of Chemical Physics}\ }\textbf {\bibinfo
  {volume} {102}},\ \bibinfo {pages} {4611} (\bibinfo {year}
  {1995}{\natexlab{b}})},\ \Eprint
  {http://arxiv.org/abs/https://doi.org/10.1063/1.469509}
  {https://doi.org/10.1063/1.469509} \BibitemShut {NoStop}%
\bibitem [{\citenamefont {Thorwart}\ \emph {et~al.}(2000)\citenamefont
  {Thorwart}, \citenamefont {Reimann},\ and\ \citenamefont
  {H\"anggi}}]{Thorwart2000}%
  \BibitemOpen
  \bibfield  {author} {\bibinfo {author} {\bibfnamefont {M.}~\bibnamefont
  {Thorwart}}, \bibinfo {author} {\bibfnamefont {P.}~\bibnamefont {Reimann}}, \
  and\ \bibinfo {author} {\bibfnamefont {P.}~\bibnamefont {H\"anggi}},\ }\href
  {\doibase 10.1103/PhysRevE.62.5808} {\bibfield  {journal} {\bibinfo
  {journal} {Phys. Rev. E}\ }\textbf {\bibinfo {volume} {62}},\ \bibinfo
  {pages} {5808} (\bibinfo {year} {2000})}\BibitemShut {NoStop}%
\bibitem [{\citenamefont {Vagov}\ \emph {et~al.}(2011)\citenamefont {Vagov},
  \citenamefont {Croitoru}, \citenamefont {Gl\"assl}, \citenamefont {Axt},\
  and\ \citenamefont {Kuhn}}]{Vagov2011}%
  \BibitemOpen
  \bibfield  {author} {\bibinfo {author} {\bibfnamefont {A.}~\bibnamefont
  {Vagov}}, \bibinfo {author} {\bibfnamefont {M.~D.}\ \bibnamefont {Croitoru}},
  \bibinfo {author} {\bibfnamefont {M.}~\bibnamefont {Gl\"assl}}, \bibinfo
  {author} {\bibfnamefont {V.~M.}\ \bibnamefont {Axt}}, \ and\ \bibinfo
  {author} {\bibfnamefont {T.}~\bibnamefont {Kuhn}},\ }\href {\doibase
  10.1103/PhysRevB.83.094303} {\bibfield  {journal} {\bibinfo  {journal} {Phys.
  Rev. B}\ }\textbf {\bibinfo {volume} {83}},\ \bibinfo {pages} {094303}
  (\bibinfo {year} {2011})}\BibitemShut {NoStop}%
\bibitem [{\citenamefont {Shao}\ and\ \citenamefont {Makri}(2002)}]{Shao2002}%
  \BibitemOpen
  \bibfield  {author} {\bibinfo {author} {\bibfnamefont {J.}~\bibnamefont
  {Shao}}\ and\ \bibinfo {author} {\bibfnamefont {N.}~\bibnamefont {Makri}},\
  }\href {\doibase 10.1063/1.1423936} {\bibfield  {journal} {\bibinfo
  {journal} {The Journal of Chemical Physics}\ }\textbf {\bibinfo {volume}
  {116}},\ \bibinfo {pages} {507} (\bibinfo {year} {2002})},\ \Eprint
  {http://arxiv.org/abs/https://doi.org/10.1063/1.1423936}
  {https://doi.org/10.1063/1.1423936} \BibitemShut {NoStop}%
\bibitem [{\citenamefont {Barth}\ \emph
  {et~al.}(2016{\natexlab{a}})\citenamefont {Barth}, \citenamefont {Vagov},\
  and\ \citenamefont {Axt}}]{Barth2016}%
  \BibitemOpen
  \bibfield  {author} {\bibinfo {author} {\bibfnamefont {A.~M.}\ \bibnamefont
  {Barth}}, \bibinfo {author} {\bibfnamefont {A.}~\bibnamefont {Vagov}}, \ and\
  \bibinfo {author} {\bibfnamefont {V.~M.}\ \bibnamefont {Axt}},\ }\href
  {\doibase 10.1103/PhysRevB.94.125439} {\bibfield  {journal} {\bibinfo
  {journal} {Phys. Rev. B}\ }\textbf {\bibinfo {volume} {94}},\ \bibinfo
  {pages} {125439} (\bibinfo {year} {2016}{\natexlab{a}})}\BibitemShut
  {NoStop}%
\bibitem [{\citenamefont {Cygorek}\ \emph {et~al.}(2017)\citenamefont
  {Cygorek}, \citenamefont {Barth}, \citenamefont {Ungar}, \citenamefont
  {Vagov},\ and\ \citenamefont {Axt}}]{Cygorek2017}%
  \BibitemOpen
  \bibfield  {author} {\bibinfo {author} {\bibfnamefont {M.}~\bibnamefont
  {Cygorek}}, \bibinfo {author} {\bibfnamefont {A.~M.}\ \bibnamefont {Barth}},
  \bibinfo {author} {\bibfnamefont {F.}~\bibnamefont {Ungar}}, \bibinfo
  {author} {\bibfnamefont {A.}~\bibnamefont {Vagov}}, \ and\ \bibinfo {author}
  {\bibfnamefont {V.~M.}\ \bibnamefont {Axt}},\ }\href {\doibase
  10.1103/PhysRevB.96.201201} {\bibfield  {journal} {\bibinfo  {journal} {Phys.
  Rev. B}\ }\textbf {\bibinfo {volume} {96}},\ \bibinfo {pages} {201201}
  (\bibinfo {year} {2017})}\BibitemShut {NoStop}%
\bibitem [{\citenamefont {Strathearn}\ \emph {et~al.}(2018)\citenamefont
  {Strathearn}, \citenamefont {Kirton}, \citenamefont {Kilda}, \citenamefont
  {Keeling},\ and\ \citenamefont {Lovett}}]{Strathearn2018}%
  \BibitemOpen
  \bibfield  {author} {\bibinfo {author} {\bibfnamefont {A.}~\bibnamefont
  {Strathearn}}, \bibinfo {author} {\bibfnamefont {P.}~\bibnamefont {Kirton}},
  \bibinfo {author} {\bibfnamefont {D.}~\bibnamefont {Kilda}}, \bibinfo
  {author} {\bibfnamefont {J.}~\bibnamefont {Keeling}}, \ and\ \bibinfo
  {author} {\bibfnamefont {B.~W.}\ \bibnamefont {Lovett}},\ }\href {\doibase
  10.1038/s41467-018-05617-3} {\bibfield  {journal} {\bibinfo  {journal}
  {Nature Communications}\ }\textbf {\bibinfo {volume} {9}},\ \bibinfo {pages}
  {3322} (\bibinfo {year} {2018})}\BibitemShut {NoStop}%
\bibitem [{\citenamefont {Cygorek}\ \emph {et~al.}(2018)\citenamefont
  {Cygorek}, \citenamefont {Ungar}, \citenamefont {Seidelmann}, \citenamefont
  {Barth}, \citenamefont {Vagov}, \citenamefont {Axt},\ and\ \citenamefont
  {Kuhn}}]{Cygorek2017b}%
  \BibitemOpen
  \bibfield  {author} {\bibinfo {author} {\bibfnamefont {M.}~\bibnamefont
  {Cygorek}}, \bibinfo {author} {\bibfnamefont {F.}~\bibnamefont {Ungar}},
  \bibinfo {author} {\bibfnamefont {T.}~\bibnamefont {Seidelmann}}, \bibinfo
  {author} {\bibfnamefont {A.~M.}\ \bibnamefont {Barth}}, \bibinfo {author}
  {\bibfnamefont {A.}~\bibnamefont {Vagov}}, \bibinfo {author} {\bibfnamefont
  {V.~M.}\ \bibnamefont {Axt}}, \ and\ \bibinfo {author} {\bibfnamefont
  {T.}~\bibnamefont {Kuhn}},\ }\href {\doibase 10.1103/PhysRevB.98.045303}
  {\bibfield  {journal} {\bibinfo  {journal} {Phys. Rev. B}\ }\textbf {\bibinfo
  {volume} {98}},\ \bibinfo {pages} {045303} (\bibinfo {year}
  {2018})}\BibitemShut {NoStop}%
\bibitem [{\citenamefont {Machnikowski}\ and\ \citenamefont
  {Jacak}(2004)}]{Machnikowski2004}%
  \BibitemOpen
  \bibfield  {author} {\bibinfo {author} {\bibfnamefont {P.}~\bibnamefont
  {Machnikowski}}\ and\ \bibinfo {author} {\bibfnamefont {L.}~\bibnamefont
  {Jacak}},\ }\href {\doibase 10.1103/PhysRevB.69.193302} {\bibfield  {journal}
  {\bibinfo  {journal} {Phys. Rev. B}\ }\textbf {\bibinfo {volume} {69}},\
  \bibinfo {pages} {193302} (\bibinfo {year} {2004})}\BibitemShut {NoStop}%
\bibitem [{\citenamefont {Louisell}(1973)}]{Louisell1973}%
  \BibitemOpen
  \bibfield  {author} {\bibinfo {author} {\bibfnamefont {W.~H.}\ \bibnamefont
  {Louisell}},\ }\href@noop {} {\emph {\bibinfo {title} {Quantum Statistical
  Properties of Radiation}}}\ (\bibinfo  {publisher} {Wiley},\ \bibinfo {year}
  {1973})\BibitemShut {NoStop}%
\bibitem [{\citenamefont {Krummheuer}\ \emph {et~al.}(2005)\citenamefont
  {Krummheuer}, \citenamefont {Axt}, \citenamefont {Kuhn}, \citenamefont
  {D'Amico},\ and\ \citenamefont {Rossi}}]{Krummheuer2005}%
  \BibitemOpen
  \bibfield  {author} {\bibinfo {author} {\bibfnamefont {B.}~\bibnamefont
  {Krummheuer}}, \bibinfo {author} {\bibfnamefont {V.~M.}\ \bibnamefont {Axt}},
  \bibinfo {author} {\bibfnamefont {T.}~\bibnamefont {Kuhn}}, \bibinfo {author}
  {\bibfnamefont {I.}~\bibnamefont {D'Amico}}, \ and\ \bibinfo {author}
  {\bibfnamefont {F.}~\bibnamefont {Rossi}},\ }\href {\doibase
  10.1103/PhysRevB.71.235329} {\bibfield  {journal} {\bibinfo  {journal} {Phys.
  Rev. B}\ }\textbf {\bibinfo {volume} {71}},\ \bibinfo {pages} {235329}
  (\bibinfo {year} {2005})}\BibitemShut {NoStop}%
\bibitem [{\citenamefont {Breuer}\ \emph {et~al.}(2009)\citenamefont {Breuer},
  \citenamefont {Laine},\ and\ \citenamefont {Piilo}}]{Breuer2009}%
  \BibitemOpen
  \bibfield  {author} {\bibinfo {author} {\bibfnamefont {H.-P.}\ \bibnamefont
  {Breuer}}, \bibinfo {author} {\bibfnamefont {E.-M.}\ \bibnamefont {Laine}}, \
  and\ \bibinfo {author} {\bibfnamefont {J.}~\bibnamefont {Piilo}},\ }\href
  {\doibase 10.1103/PhysRevLett.103.210401} {\bibfield  {journal} {\bibinfo
  {journal} {Phys. Rev. Lett.}\ }\textbf {\bibinfo {volume} {103}},\ \bibinfo
  {pages} {210401} (\bibinfo {year} {2009})}\BibitemShut {NoStop}%
\bibitem [{\citenamefont {Reiter}(2017)}]{Reiter2017}%
  \BibitemOpen
  \bibfield  {author} {\bibinfo {author} {\bibfnamefont {D.~E.}\ \bibnamefont
  {Reiter}},\ }\href {\doibase 10.1103/PhysRevB.95.125308} {\bibfield
  {journal} {\bibinfo  {journal} {Phys. Rev. B}\ }\textbf {\bibinfo {volume}
  {95}},\ \bibinfo {pages} {125308} (\bibinfo {year} {2017})}\BibitemShut
  {NoStop}%
\bibitem [{\citenamefont {Iles-Smith}\ \emph
  {et~al.}(2017{\natexlab{a}})\citenamefont {Iles-Smith}, \citenamefont
  {McCutcheon}, \citenamefont {M\o{}rk},\ and\ \citenamefont
  {Nazir}}]{Iles-Smith2017}%
  \BibitemOpen
  \bibfield  {author} {\bibinfo {author} {\bibfnamefont {J.}~\bibnamefont
  {Iles-Smith}}, \bibinfo {author} {\bibfnamefont {D.~P.~S.}\ \bibnamefont
  {McCutcheon}}, \bibinfo {author} {\bibfnamefont {J.}~\bibnamefont {M\o{}rk}},
  \ and\ \bibinfo {author} {\bibfnamefont {A.}~\bibnamefont {Nazir}},\ }\href
  {\doibase 10.1103/PhysRevB.95.201305} {\bibfield  {journal} {\bibinfo
  {journal} {Phys. Rev. B}\ }\textbf {\bibinfo {volume} {95}},\ \bibinfo
  {pages} {201305} (\bibinfo {year} {2017}{\natexlab{a}})}\BibitemShut
  {NoStop}%
\bibitem [{\citenamefont {Iles-Smith}\ \emph
  {et~al.}(2017{\natexlab{b}})\citenamefont {Iles-Smith}, \citenamefont
  {McCutcheon}, \citenamefont {Nazir},\ and\ \citenamefont
  {M{\o}rk}}]{Iles-Smith2017b}%
  \BibitemOpen
  \bibfield  {author} {\bibinfo {author} {\bibfnamefont {J.}~\bibnamefont
  {Iles-Smith}}, \bibinfo {author} {\bibfnamefont {D.~P.~S.}\ \bibnamefont
  {McCutcheon}}, \bibinfo {author} {\bibfnamefont {A.}~\bibnamefont {Nazir}}, \
  and\ \bibinfo {author} {\bibfnamefont {J.}~\bibnamefont {M{\o}rk}},\ }\href
  {http://dx.doi.org/10.1038/nphoton.2017.101} {\bibfield  {journal} {\bibinfo
  {journal} {Nature Photonics}\ }\textbf {\bibinfo {volume} {11}},\ \bibinfo
  {pages} {521} (\bibinfo {year} {2017}{\natexlab{b}})}\BibitemShut {NoStop}%
\bibitem [{\citenamefont {Mollow}(1969)}]{Mollow1969}%
  \BibitemOpen
  \bibfield  {author} {\bibinfo {author} {\bibfnamefont {B.~R.}\ \bibnamefont
  {Mollow}},\ }\href {\doibase 10.1103/PhysRev.188.1969} {\bibfield  {journal}
  {\bibinfo  {journal} {Phys. Rev.}\ }\textbf {\bibinfo {volume} {188}},\
  \bibinfo {pages} {1969} (\bibinfo {year} {1969})}\BibitemShut {NoStop}%
\bibitem [{\citenamefont {Kr{\"u}gel}\ \emph {et~al.}(2005)\citenamefont
  {Kr{\"u}gel}, \citenamefont {Axt}, \citenamefont {Kuhn}, \citenamefont
  {Machnikowski},\ and\ \citenamefont {Vagov}}]{Kruegel2005}%
  \BibitemOpen
  \bibfield  {author} {\bibinfo {author} {\bibfnamefont {A.}~\bibnamefont
  {Kr{\"u}gel}}, \bibinfo {author} {\bibfnamefont {V.}~\bibnamefont {Axt}},
  \bibinfo {author} {\bibfnamefont {T.}~\bibnamefont {Kuhn}}, \bibinfo {author}
  {\bibfnamefont {P.}~\bibnamefont {Machnikowski}}, \ and\ \bibinfo {author}
  {\bibfnamefont {A.}~\bibnamefont {Vagov}},\ }\href {\doibase
  10.1007/s00340-005-1984-1} {\bibfield  {journal} {\bibinfo  {journal}
  {Applied Physics B}\ }\textbf {\bibinfo {volume} {81}},\ \bibinfo {pages}
  {897} (\bibinfo {year} {2005})}\BibitemShut {NoStop}%
\bibitem [{\citenamefont {Nazir}(2008)}]{Nazir2008}%
  \BibitemOpen
  \bibfield  {author} {\bibinfo {author} {\bibfnamefont {A.}~\bibnamefont
  {Nazir}},\ }\href {\doibase 10.1103/PhysRevB.78.153309} {\bibfield  {journal}
  {\bibinfo  {journal} {Phys. Rev. B}\ }\textbf {\bibinfo {volume} {78}},\
  \bibinfo {pages} {153309} (\bibinfo {year} {2008})}\BibitemShut {NoStop}%
\bibitem [{\citenamefont {Ramsay}\ \emph
  {et~al.}(2010{\natexlab{b}})\citenamefont {Ramsay}, \citenamefont {Godden},
  \citenamefont {Boyle}, \citenamefont {Gauger}, \citenamefont {Nazir},
  \citenamefont {Lovett}, \citenamefont {Fox},\ and\ \citenamefont
  {Skolnick}}]{Ramsay2010b}%
  \BibitemOpen
  \bibfield  {author} {\bibinfo {author} {\bibfnamefont {A.~J.}\ \bibnamefont
  {Ramsay}}, \bibinfo {author} {\bibfnamefont {T.~M.}\ \bibnamefont {Godden}},
  \bibinfo {author} {\bibfnamefont {S.~J.}\ \bibnamefont {Boyle}}, \bibinfo
  {author} {\bibfnamefont {E.~M.}\ \bibnamefont {Gauger}}, \bibinfo {author}
  {\bibfnamefont {A.}~\bibnamefont {Nazir}}, \bibinfo {author} {\bibfnamefont
  {B.~W.}\ \bibnamefont {Lovett}}, \bibinfo {author} {\bibfnamefont {A.~M.}\
  \bibnamefont {Fox}}, \ and\ \bibinfo {author} {\bibfnamefont {M.~S.}\
  \bibnamefont {Skolnick}},\ }\href {\doibase 10.1103/PhysRevLett.105.177402}
  {\bibfield  {journal} {\bibinfo  {journal} {Phys. Rev. Lett.}\ }\textbf
  {\bibinfo {volume} {105}},\ \bibinfo {pages} {177402} (\bibinfo {year}
  {2010}{\natexlab{b}})}\BibitemShut {NoStop}%
\bibitem [{\citenamefont {Wei}\ \emph {et~al.}(2014)\citenamefont {Wei},
  \citenamefont {He}, \citenamefont {He}, \citenamefont {Lu}, \citenamefont
  {Pan}, \citenamefont {Schneider}, \citenamefont {Kamp}, \citenamefont
  {H\"ofling}, \citenamefont {McCutcheon},\ and\ \citenamefont
  {Nazir}}]{Wei2014}%
  \BibitemOpen
  \bibfield  {author} {\bibinfo {author} {\bibfnamefont {Y.-J.}\ \bibnamefont
  {Wei}}, \bibinfo {author} {\bibfnamefont {Y.}~\bibnamefont {He}}, \bibinfo
  {author} {\bibfnamefont {Y.-M.}\ \bibnamefont {He}}, \bibinfo {author}
  {\bibfnamefont {C.-Y.}\ \bibnamefont {Lu}}, \bibinfo {author} {\bibfnamefont
  {J.-W.}\ \bibnamefont {Pan}}, \bibinfo {author} {\bibfnamefont
  {C.}~\bibnamefont {Schneider}}, \bibinfo {author} {\bibfnamefont
  {M.}~\bibnamefont {Kamp}}, \bibinfo {author} {\bibfnamefont {S.}~\bibnamefont
  {H\"ofling}}, \bibinfo {author} {\bibfnamefont {D.~P.~S.}\ \bibnamefont
  {McCutcheon}}, \ and\ \bibinfo {author} {\bibfnamefont {A.}~\bibnamefont
  {Nazir}},\ }\href {\doibase 10.1103/PhysRevLett.113.097401} {\bibfield
  {journal} {\bibinfo  {journal} {Phys. Rev. Lett.}\ }\textbf {\bibinfo
  {volume} {113}},\ \bibinfo {pages} {097401} (\bibinfo {year}
  {2014})}\BibitemShut {NoStop}%
\bibitem [{\citenamefont {Gl\"assl}\ and\ \citenamefont
  {Axt}(2012)}]{Glaessl2012}%
  \BibitemOpen
  \bibfield  {author} {\bibinfo {author} {\bibfnamefont {M.}~\bibnamefont
  {Gl\"assl}}\ and\ \bibinfo {author} {\bibfnamefont {V.~M.}\ \bibnamefont
  {Axt}},\ }\href {\doibase 10.1103/PhysRevB.86.245306} {\bibfield  {journal}
  {\bibinfo  {journal} {Phys. Rev. B}\ }\textbf {\bibinfo {volume} {86}},\
  \bibinfo {pages} {245306} (\bibinfo {year} {2012})}\BibitemShut {NoStop}%
\bibitem [{\citenamefont {Barth}\ \emph
  {et~al.}(2016{\natexlab{b}})\citenamefont {Barth}, \citenamefont {L\"uker},
  \citenamefont {Vagov}, \citenamefont {Reiter}, \citenamefont {Kuhn},\ and\
  \citenamefont {Axt}}]{Barth2016b}%
  \BibitemOpen
  \bibfield  {author} {\bibinfo {author} {\bibfnamefont {A.~M.}\ \bibnamefont
  {Barth}}, \bibinfo {author} {\bibfnamefont {S.}~\bibnamefont {L\"uker}},
  \bibinfo {author} {\bibfnamefont {A.}~\bibnamefont {Vagov}}, \bibinfo
  {author} {\bibfnamefont {D.~E.}\ \bibnamefont {Reiter}}, \bibinfo {author}
  {\bibfnamefont {T.}~\bibnamefont {Kuhn}}, \ and\ \bibinfo {author}
  {\bibfnamefont {V.~M.}\ \bibnamefont {Axt}},\ }\href {\doibase
  10.1103/PhysRevB.94.045306} {\bibfield  {journal} {\bibinfo  {journal} {Phys.
  Rev. B}\ }\textbf {\bibinfo {volume} {94}},\ \bibinfo {pages} {045306}
  (\bibinfo {year} {2016}{\natexlab{b}})}\BibitemShut {NoStop}%
\bibitem [{\citenamefont {Nahri}\ \emph {et~al.}(2017)\citenamefont {Nahri},
  \citenamefont {Mathkoor},\ and\ \citenamefont {Ooi}}]{Nahri2017}%
  \BibitemOpen
  \bibfield  {author} {\bibinfo {author} {\bibfnamefont {D.~G.}\ \bibnamefont
  {Nahri}}, \bibinfo {author} {\bibfnamefont {F.~H.~A.}\ \bibnamefont
  {Mathkoor}}, \ and\ \bibinfo {author} {\bibfnamefont {C.~H.~R.}\ \bibnamefont
  {Ooi}},\ }\href {http://stacks.iop.org/0953-8984/29/i=5/a=055701} {\bibfield
  {journal} {\bibinfo  {journal} {Journal of Physics: Condensed Matter}\
  }\textbf {\bibinfo {volume} {29}},\ \bibinfo {pages} {055701} (\bibinfo
  {year} {2017})}\BibitemShut {NoStop}%
\bibitem [{\citenamefont {Chakraborty}\ and\ \citenamefont
  {Sensarma}(2018)}]{Chakraborty2018}%
  \BibitemOpen
  \bibfield  {author} {\bibinfo {author} {\bibfnamefont {A.}~\bibnamefont
  {Chakraborty}}\ and\ \bibinfo {author} {\bibfnamefont {R.}~\bibnamefont
  {Sensarma}},\ }\href {\doibase 10.1103/PhysRevB.97.104306} {\bibfield
  {journal} {\bibinfo  {journal} {Phys. Rev. B}\ }\textbf {\bibinfo {volume}
  {97}},\ \bibinfo {pages} {104306} (\bibinfo {year} {2018})}\BibitemShut
  {NoStop}%
\end{thebibliography}%

\end{document}